\documentclass[aps,prl,reprint,amsmath,amssymb,superscriptaddress,nofootinbib,noeprint]{revtex4-2}
\pdfoutput=1

\usepackage{txfonts}
\usepackage{graphicx}% Include figure files
\usepackage{dcolumn}% Align table columns on decimal point
\usepackage{bm}% bold math
\usepackage[version=3]{mhchem}
\usepackage{mathtools}

\PassOptionsToPackage{hyphens}{url}

\usepackage{float}
\usepackage{lineno}
\usepackage{xcolor}
\usepackage{graphicx}
\usepackage{amsmath,amssymb}
\usepackage{upgreek}
\usepackage{graphicx}
\usepackage{braket}
\usepackage{units}
\usepackage{times}
\usepackage{placeins}
\usepackage{soul}
\DeclareMathOperator{\Tr}{Tr}

\usepackage{xcolor}
\definecolor{mypink1}{rgb}{0.858, 0.188, 0.478}
\definecolor{myColor}{rgb}{0.02,0.12,0.3}
\definecolor{myciteColor}{rgb}{0.39,0.7,0.89}
\definecolor{myciteColor}{rgb}{0.0,0.5,0.23}
\usepackage[colorlinks=true,citecolor=myciteColor,linkcolor=myciteColor,urlcolor=myciteColor]{hyp
erref}

\newcommand{\RNum}[1]{\uppercase\expandafter{\romannumeral #1\relax.}}
\makeatletter
\def\maketitle{
\@author@finish
\title@column\titleblock@produce
\suppressfloats[t]}

\newcommand{\F}{\mathrm{F}}
\newcommand{\Q}{\mathrm{Q}}
\newcommand{\B}{\mathrm{B}}
\newcommand{\R}{\mathrm{R}}

\newcommand{\EF}{E_\mathrm{F}}
\newcommand{\Ep}{E_\mathrm{p}}
\newcommand{\kF}{k_\mathrm{F}}
\newcommand{\kB}{k_\mathrm{B}}
\newcommand{\kFa}{k_\mathrm{F}a}

\newcommand{\OmegaR}{\Omega_\mathrm{R}}

\newcommand{\mbf}{\mathbf}
\newcommand{\Z}{\mathcal{Z}}

\begin{document}

\title{The strongly driven Fermi polaron}

 \author{Franklin J. Vivanco$^*$}
 \affiliation{Department of Physics, Yale University, New Haven, Connecticut 06520, USA}
 \author{Alexander Schuckert$^{*\dag}$}
 \affiliation{Joint Quantum Institute and Joint Center for Quantum Information and Computer Science, University of Maryland and NIST, College Park, Maryland 20742, USA}
 \author{Songtao Huang$^{*\dag}$}
 \affiliation{Department of Physics, Yale University, New Haven, Connecticut 06520, USA}
 \author{Grant L. Schumacher}
 \affiliation{Department of Physics, Yale University, New Haven, Connecticut 06520, USA}
 \author{Gabriel G. T. Assump{\c{c}}{\~a}o}
 \affiliation{Department of Physics, Yale University, New Haven, Connecticut 06520, USA}
 \author{Yunpeng Ji}
 \affiliation{Department of Physics, Yale University, New Haven, Connecticut 06520, USA}
 \author{Jianyi Chen}
 \affiliation{Department of Physics, Yale University, New Haven, Connecticut 06520, USA}
 \author{Michael Knap}
 \affiliation{Technical University of Munich, TUM School of Natural Sciences, Physics Department, 85748 Garching, Germany}
 \affiliation{Munich Center for Quantum Science and Technology (MCQST), Schellingstr. 4, 80799 M{\"u}nchen, Germany}
 \author{Nir Navon}
 \affiliation{Department of Physics, Yale University, New Haven, Connecticut 06520, USA}
 \affiliation{Yale Quantum Institute, Yale University, New Haven, Connecticut 06520, USA}

\def\thefootnote{*}\footnotetext{These authors contributed equally to this work}
\def\thefootnote{$\dagger$}\footnotetext{Corresponding authors: aschu@umd.edu, songtao.huang@yale.edu}
\date{\today}

\begin{abstract}
Quasiparticles are emergent excitations of matter that underlie much of our understanding of quantum many-body systems~\cite{Landau1956}. Therefore, the prospect of manipulating their properties with external fields -- or even destroying them -- has both fundamental and practical implications~\cite{basov2017towards,de2021colloquium,bloch2022strongly}.  
However, in solid-state
materials it is often challenging to understand how quasiparticles are modified by external fields owing to their complex interplay with other collective excitations, such as phonons~\cite{basov2017towards}. Here, we take advantage of the clean setting of homogeneous quantum gases~\cite{navon2021quantum} and fast radio-frequency control~\cite{kohstall2012metastability, cetina2016ultrafast} to manipulate Fermi polarons -- quasiparticles formed by impurities interacting with a non-interacting Fermi gas~\cite{schirotzek2009observation, Nascimbene2009,radzihovsky2010imbalanced,chevy2010ultracold,Koschorreck2012,Zhang2012,Oppong2019} -- from weak to ultrastrong drives. Exploiting two internal states of the impurity species, we develop a steady-state spectroscopy, from which we extract the energy of the driven polaron. We measure the decay rate and the quasiparticle residue of the driven polaron from the Rabi oscillations between the two internal states. At large drive strengths, the so-extracted quasiparticle residue exceeds unity, raising intriguing questions on the relationship between the Rabi oscillations and the impurity's spectral functions. Our experiment establishes the driven Fermi polaron as a promising platform for studying controllable quasiparticles in strongly driven quantum matter.
\end{abstract}

\maketitle

\begin{figure*}[bt]
  \centering
  \includegraphics[width=2\columnwidth]{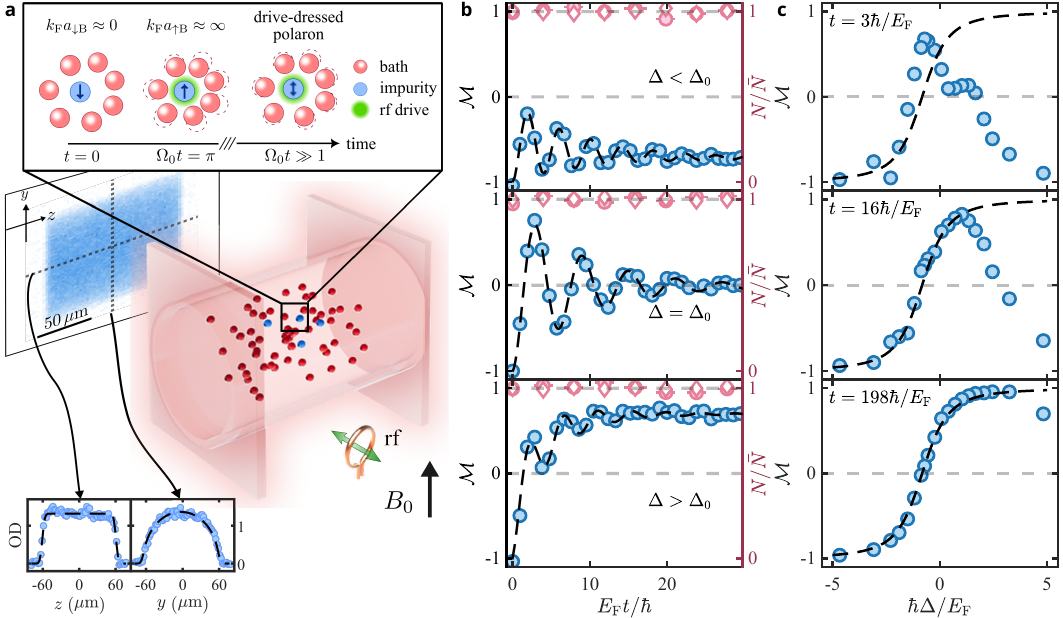}
  
  \caption{\textbf{The strongly driven Fermi impurity.  
  } (\textbf{a}) A uniform Fermi gas of $^6$Li is trapped in a cylindrical optical box. The red (blue) dots inside the box are bath (impurity) atoms. The image is a typical \emph{in situ} optical density (OD) image of the impurity species (averaged over 14 realizations). Lower insets: Central cuts of the OD image. The black dashed lines are fits to the density profiles.
  Upper inset: cartoon of a typical Rabi oscillation experiment and subsequent dressing of the Fermi impurity by both its interactions with the bath and its coupling to the radio-frequency (rf) field, \emph{i.e. the drive-dressed polaron}. 
  The impurity is initialized in state $\ket{\downarrow}$, which interacts weakly with $\ket{\B}$. After a half-cycle, it is predominantly in $\ket{\uparrow}$, interacting strongly with $\ket{\B}$. At long times, it reaches a steady state that is a mixture of states $\ket{\uparrow}$ and $\ket{\downarrow}$ (see text). (\textbf{b}) Rabi oscillations of the magnetization $\mathcal{M}$ for three detunings $\Delta$ from the bare transition between $\ket{\downarrow}$ and $\ket{\uparrow}$ (full blue circles with left axis), $\Delta/\Delta_0=-0.3,\,1,\,2.4$ (from top to bottom); $\hbar/\EF\approx25\,\mu$s. The pink circles (diamonds) are the total impurity (bath) atom numbers $N=N_\uparrow+N_\downarrow$ ($N=N_\B$), normalized by the atom number averaged over the whole time series $\Bar{N}$ (right axis). The vertical error bars are the standard error of the mean (s.e.m), which is smaller than the marker size; the horizontal error bars correspond to the standard deviation of $\EF t/\hbar$ in each time bin.  (\textbf{c}) Magnetization as a function of detuning with pulse time $t$ increasing from top to bottom.  In (\textbf{b}) and (\textbf{c}), the Rabi frequency is $\Omega_0/(2\pi)\approx\unit[8]{kHz}$, and the black dashed lines are guides to the eye.\label{FIG:1}}
\end{figure*}

Often, the low-energy physics of quantum many-body systems can be understood in terms of quasiparticle excitations. This description has been successful in explaining the thermodynamic and near-equilibrium transport properties of a wide range of materials~\cite{Landau1956}. Methods to modify, and even controllably destroy, these quasiparticles have long been sought after~\cite{basov2017towards,de2021colloquium}. Such tuning capability could allow modifying the thermodynamics of a system, and potentially offer a route towards realizing strongly correlated systems without well-defined quasiparticles~\cite{P_Coleman_2001,Hartnoll:2018xxg}.

Fermi polarons, impurities that interact with a non-interacting Fermi gas, have attracted considerable interest because they constitute one of the simplest quantum many-body settings for studying both in- and out-of-equilibrium properties in correlated systems~\cite{schirotzek2009observation,kohstall2012metastability,massignan2014polarons,cetina2016ultrafast,scazza2022repulsive}. Both attractive and repulsive Fermi polarons have been realized experimentally with ultracold atoms~\cite{schirotzek2009observation,kohstall2012metastability,Zhang2012,Scazza2017,yan2019boiling,Oppong2019} and with semiconductor heterostructures~\cite{sidler_polaron}. Furthermore, Rabi oscillations of impurities with internal states that interact differently with the bath have been used to probe the properties of these polarons, though the interpretation of these experiments remains challenging~\cite{kohstall2012metastability,PhysRevB.94.184303,Scazza2017,Adlong2020,Hu2022,wasak2022decoherence}. Harnessing the non-equilibrium dynamics of driven Fermi polarons to manipulate their quasiparticle properties has thus remained a major challenge to both experiment and theory.

Here, we investigate Fermi impurities embedded in a homogeneous atomic Fermi gas and driven with an external radio-frequency (rf) field as a platform for realizing quasiparticles with tunable properties (Fig.~\ref{FIG:1}a). The impurities have two internal states: one of them essentially does not interact with the Fermi gas, whereas the other interacts unitarily with it. We drive the impurity by coupling the two internal states with the rf field and probe the impurities dressed by both their interactions with the bath and their coupling to the rf field; we measure their quasiparticle properties -- energy, decay rate, and residue -- as a function of the drive strength.

Our experiment begins with a spatially uniform quantum gas of $^6$Li atoms confined in an optical box~\cite{mukherjee2017homogeneous,navon2021quantum} (see Fig.~\ref{FIG:1}a). The gas is in a highly imbalanced mixture of the internal states $\ket{\downarrow}$ and $\ket{\B}$, where $\ket{\downarrow}$ is one of the two internal states of the impurity and $\ket{\B}$ is the bath state which forms the Fermi gas (see Methods). The bath has a Fermi energy $\EF/\hbar\approx 2\pi\times\unit[6]{kHz}$ and a temperature $T/T_\F=0.25(2)$, where $T_\F$ is the Fermi temperature of the bath and $\hbar$ is the reduced Planck's constant. The magnetic field is set to the Feshbach resonance $B_0$ of the $\ket{\uparrow}$-$\ket{\B}$ mixture, while $\ket{\downarrow}$ and $\ket{\B}$ are weakly interacting; \emph{i.e.} $1/k_\F a_{\uparrow\B}=0$ and $k_\F a_{\downarrow\B}\approx 0.16$, where $\ket{\uparrow}$ is the second internal state of the impurity, $k_\F$ is the Fermi wavevector of the bath, and $a_{\uparrow\B}$ (resp. $a_{\downarrow\B}$) is the s-wave scattering length between states $\ket{\uparrow}$ (resp. $\ket{\downarrow}$) and $\ket{\B}$. An \emph{in situ} optical density image of the impurity species $\ket{\downarrow}$ is shown in Fig.~\ref{FIG:1}a, along with two central cuts.

After initialization, an rf field of Rabi frequency $\Omega_0$ and detuning $\Delta$ (relative to the bare $\ket{\downarrow}$-$\ket{\uparrow}$ transition) is turned on for a duration $t$. We then measure the magnetization of the impurities $\mathcal{M}\equiv (N_\uparrow-N_\downarrow)/(N_\uparrow+N_\downarrow)$ as a function of $t$, where $N_\uparrow$ ($N_\downarrow$) is the population of the impurities in state $\ket{\uparrow}$ ($\ket{\downarrow}$). In Fig.~\ref{FIG:1}b, we show typical measurements of the dynamics of the magnetization for different detunings.
For all detunings we observe damped Rabi oscillations, which reach a detuning-dependent steady-state value at long times. The dependence of the magnetization on the detuning is visualized in Fig.~\ref{FIG:1}c, where we show the magnetization as a function of the detuning for various evolution times. At short times, we see a sharp peak with a broad shoulder reminiscent of the linear-response spectrum at unitarity~\cite{footnote1,cetina2016ultrafast,kohstall2012metastability,Scazza2017}. However, at longer times we find a strong deviation from linear-response-type behavior; in fact, $\mathcal{M}$ converges to a monotonously increasing function of $\Delta$ in the steady state (black dashed lines in Fig.~\ref{FIG:1}c). The steady-state magnetization spectrum  $\mathcal{M}(\Delta)$ vanishes at a detuning $\Delta_0$, the \emph{zero crossing}.

\begin{figure*}[bt]
  \centering
  \includegraphics[width=2\columnwidth]{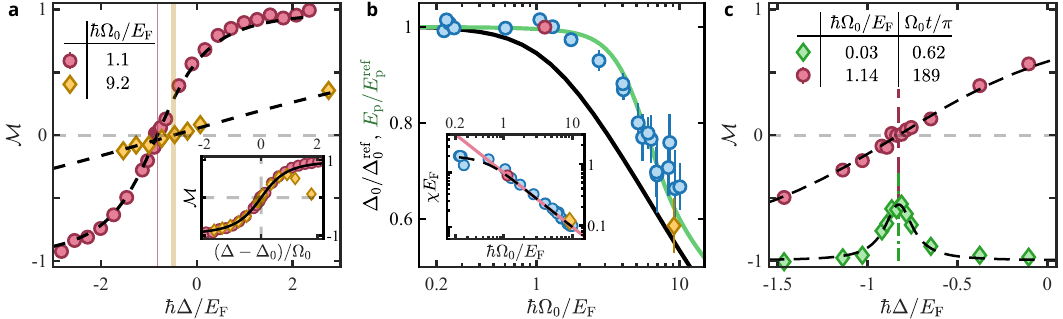}
  \caption{\textbf{Many-body steady-state spectroscopy}. (\textbf{a}) Steady-state spectra measured for two Rabi frequencies. The black dashed lines are fits to Eq.~\eqref{eq:M}.
  The red and orange vertical bands mark the zero crossing $\Delta_0$ including the fitting uncertainty. Inset: Scaling collapse of the steady-state spectra. The black solid line corresponds to Eq.~\eqref{eq:M}.
   (\textbf{b}) Normalized zero crossing versus drive strength. The blue, red, and orange points are the experimental data for $\Delta_0$ (see Methods); the red and orange points correspond to the spectra shown in (\textbf{a}). The error bars are the uncertainties of the fits. The black (green) solid line is the theoretical prediction for the zero crossing $\Delta_0$ (drive-dependent polaron energy $E_\mathrm{p}$), including the effect of finite temperature and corrections due to the non-zero $\kFa_{\downarrow\B}$ (see discussion in main text for the difference between $E_\mathrm{p}$ and $\Delta_0$). The normalization is chosen to be the corresponding weak-drive limits; specifically, $\hbar\Delta_0^{\mathrm{ref}}/\EF=-0.83$ ($-0.67$) for the experimental data (theory), and $\Ep^{\mathrm{ref}}/\EF=-0.75$ for the theory. The data normalized to $\EF$ is shown in the SM~\cite{supp}. Inset: Susceptibility $\chi$ versus drive strength. The red solid line is the susceptibility of the ground state Eq.~\eqref{eq:M}. The black dashed line is a fit to the $T>0$ generalization of the model in Eq.~\eqref{eq:M} (see text).
   (\textbf{c}) Comparison of linear-response injection spectroscopy (green diamonds) 
   and steady-state spectroscopy (red circles). The black dashed lines are a fit of the steady-state spectrum (resp. linear-response spectrum) to Eq.~\eqref{eq:M} (resp. to a Lorentzian function). The red dashed line (green dash-dotted line) is the fitted zero crossing (linear-response peak position). Error bars in (\textbf{a}) and (\textbf{c}) are the s.e.m of the measurements and are smaller than the marker size.}
  \label{FIG:2}
\end{figure*}

We now study how $\mathcal{M}(\Delta)$ varies with the drive strength $\hbar\Omega_0/E_\mathrm{F}$. In Fig.~\ref{FIG:2}a, we show steady-state spectra for $\hbar\Omega_0/E_\mathrm{F}=1.1$ and $9.2$. While the zero crossing and the typical width of the spectra depend on $\hbar\Omega_0/\EF$, we see in the inset of Fig.~\ref{FIG:2}a that the spectra collapse onto a universal curve when the detuning is rescaled to $(\Delta-\Delta_0)/\Omega_0$, where $\Delta_0$ depends on $\Omega_0$.

A first description of this behavior is provided by the ground state of an effective spin $1/2$ coupled to a field of Rabi frequency $\tilde{\Omega}$ and detuning $\tilde \Delta$ from its resonance (see SM~\cite{supp}): 
\begin{equation}
	\mathcal{M}_0= \frac{\tilde \Delta}{\sqrt{\tilde{\Omega}^2+\tilde \Delta^2}}.\label{eq:M}
\end{equation}
Our measured magnetization is indeed well described by this model when setting $\tilde{\Delta}=\Delta-\Delta_0$ and $\tilde{\Omega}=\Omega_0$ (black dashed line in Fig.~\ref{FIG:2}a). This model captures the rescaling of the spectrum, with $\mathcal{M}_0= f(\tilde \Delta/\Omega_0)$ and $f(x)=x/\sqrt{1+x^2}$. 

This universality suggests that our protocol realizes a novel \emph{steady-state spectroscopy}, \emph{i.e.} that the response of the many-body system is captured by a small number of (drive-dependent) parameters (in this case, $\Delta_0$ and the response's typical width). Interestingly, $\Delta_0$ is shifting closer to zero with increasing drive strength (see the vertical color bands in Fig.~\ref{FIG:2}a). 
To quantify this effect, we show the extracted $\Delta_0$ over nearly two orders of magnitude of drive strengths in Fig.~\ref{FIG:2}b. We also extract the typical width of the spectrum in the inset of Fig.~\ref{FIG:2}b, which we characterize by the on-resonance susceptibility $\chi\equiv(\partial \mathcal{M}/\partial \Delta)|_{\Delta=\Delta_0}$. The behavior of $\chi(\Omega_0)$ is well explained with the $T>0$  generalization of the model Eq.~(\ref{eq:M})~(see SM~\cite{supp}): for such a system, the finite-temperature magnetization is given by $\mathcal{M}=\mathcal{M}_0\tanh\left(\sqrt{(\hbar \Omega_0)^2+(\hbar\tilde \Delta)^2}/(2k_\B T_\text{spin})\right)$. A fit of the measured $\chi$ to this model yields a spin (\emph{i.e.} internal-state) temperature for the impurity of $T_\text{spin}=0.25(1)T_\F$ (see black dashed line in the inset of Fig.~\ref{FIG:2}b). Because the finite-temperature correction becomes negligible when $\hbar\Omega_0\gtrsim 2k_\mathrm{B} T_\text{spin}$, the impurity is effectively in its internal ground state for $\hbar\Omega_0\gtrsim E_\mathrm{F}$ (see black dashed line and red solid line in the inset of Fig.~\ref{FIG:2}b).

The fact that the spin temperature $T_\text{spin}$ matches the bath temperature $T$ indicates that the internal degrees of freedom of the driven polaron have thermalized with the fermionic bath. Two additional observations support the fact that our system consisting of the impurities plus the bath obeys closed-system dynamics. First, the coherence time of the Rabi oscillations in the absence of the bath exceeds, by about two orders of magnitude, the typical duration of our experiments in the presence of the bath. Thus, the decay is not due to dephasing coming from, for instance, magnetic field noise or inhomogeneity (see SM~\cite{supp}). Secondly, the impurity and bath atom numbers are constant during the dynamics (see pink symbols in Fig.~\ref{FIG:1}b).

A nontrivial shift of the zero crossing, $\Delta_0\neq 0$, is a prime indication of interactions between the impurity and the Fermi gas. To gain insight into $\Delta_0$, we measure the rf linear-response injection spectrum, \emph{i.e.} the fraction of impurities transferred from $\ket{\downarrow}$ into $\ket{\uparrow}$ using a weak rf pulse ($\hbar\Omega_0\ll E_\mathrm{F}$), with a small pulse area ($\Omega_0 t\ll 2\pi$). A typical measurement is shown as green diamonds in Fig.~\ref{FIG:2}c. We find that $\Delta_0$ matches the location of the peak $\Ep$ of the linear-response spectrum (\emph{i.e.} the energy of the attractive polaron~\cite{yan2019boiling}), even for a steady-state spectrum taken at a moderate rf power $\hbar\Omega_0\approx E_\mathrm{F}$ (red points in Fig.~\ref{FIG:2}c). We thus conclude that the plateau at low $\hbar\Omega_0/E_\mathrm{F}$ in Fig.~\ref{FIG:2}b matches the linear-response peak position. However, with increasing drive strength $\hbar\Omega_0/E_\mathrm{F}\gtrsim 1$, the zero crossing smoothly departs from its low-$\Omega_0$ plateau, towards zero.

We compare our measurements with the predictions of a diagrammatic theory based on the non-self-consistent $T$ matrix (see Methods and SM~\cite{supp}). We directly include the drive in the $T$ matrix~\cite{Hu2022} to describe the drive-induced changes of the scattering properties of the impurities. We also include corrections due to the nonzero $a_{\downarrow\B}$ (see Methods). In Fig.~\ref{FIG:2}b, we show as a black line the theoretical prediction for $\Delta_0(\Omega_0)$, calculated from the spectral functions (see Methods and SM~\cite{supp}); the theoretical $\Delta_0$ is normalized to its weak-drive limit $\Delta_0^\textrm{ref}\equiv\Delta_0(\Omega_0\rightarrow 0)$. We find a behavior that reproduces qualitatively the experimental data. Interestingly, the (drive-dressed) polaron energy $E_\textrm{p}$, defined as the solution of $E_\textrm{p}=\mathrm{Re}\,\Sigma(E_\textrm{p})$ (where $\Sigma$ is the zero-momentum self-energy of the impurity evaluated at the Rabi frequency $\Omega_0$ and detuning $\Delta_0$), is in remarkable agreement with the experimental data, when normalized to the weak-drive limit $E_\textrm{p}^\textrm{ref}$ (see the green solid line in Fig.~\ref{FIG:2}b and caption)~\cite{footnote}. This observation suggests that $\Delta_0$ may be the dressed polaron energy at all drive strengths, and motivates the development of more refined methods to calculate $\Delta_0$.

\begin{figure}[bt]
  \centering
  \includegraphics[width=1\columnwidth]{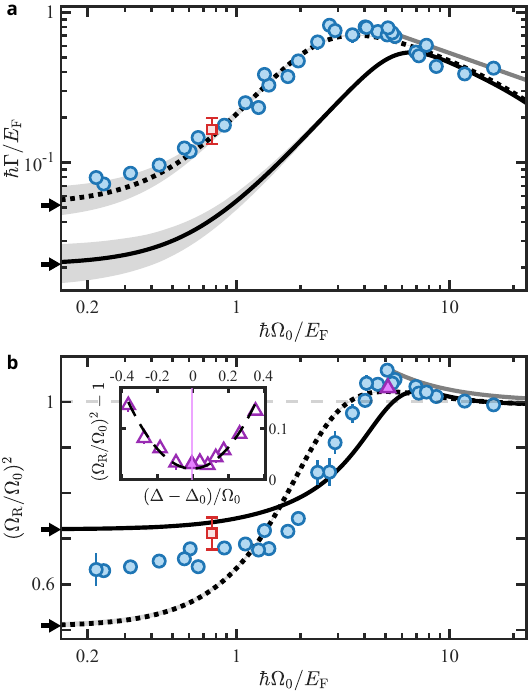}
  \caption{\textbf{Dynamical properties of the driven Fermi polaron.} (\textbf{a}) Decay rate $\Gamma$, and (\textbf{b}) the renormalized Rabi frequency $\OmegaR$ extracted from the Rabi oscillations on resonance $\Delta=\Delta_0$ (see the middle panel in Fig.~\ref{FIG:1}b for an example). The red squares are the measurements of Ref.~\cite{Scazza2017} and the purple triangle in (\textbf{b}) refers to the measurement in the inset.
  The black solid (dotted) lines are the theoretical predictions using the in-medium (two-body) $T$ matrix with error bands evaluated from the experimental uncertainty on the temperature. Arrows indicate theoretical predictions at $\Omega_0=0$. The gray solid line is the analytical result in the limit of $\hbar \Omega_0/\EF\gg 1$ (assuming $k_\F a_{\downarrow\B}=0$ and $T=0$). The error bars are the uncertainties of the fits. Inset of panel (\textbf{b}): $(\OmegaR/\Omega_0)^2-1$ as a function of the detuning relative to the zero crossing. The solid point refers to the measurement in the main figure. The dashed line is a quadratic fit. The vertical solid line marks the center of the fit, $\Delta/\Delta_0=1.03(5)$. Except for this inset, $\Omega_\mathrm{R}$ is measured at $\Delta=\Delta_0$.}  
  \label{FIG:3}
\end{figure}

We now study the pre-steady-state dynamics, focusing on the resonant case $\Delta=\Delta_0$. The magnetization undergoes damped oscillations towards $\mathcal{M}=0$ (see the middle panel of Fig.~\ref{FIG:1}b). We find that the data is well fitted by $\mathcal{M}(t)=-\cos(\Omega_\text{R}t)\exp(- \Gamma t/2)$, from which we extract a renormalized Rabi frequency $\Omega_\mathrm{R}$ and a decay rate $\Gamma$ (see Methods).

These quantities have been connected to the equilibrium properties of the polaron~\cite{kohstall2012metastability,Adlong2020}. By assuming that the bath is not perturbed by the impurities, the dynamics of the Rabi oscillations at early times can be approximately captured by an equilibrium correlation function that is given at low temperature by the impurity's zero-momentum spectral function in the presence of the drive (see SM~\cite{supp}). This approach leads to the following expressions for the zero-momentum quasiparticle decay rate $\Gamma$ and the renormalized Rabi frequency $\Omega_\mathrm{R}$: $\hbar\Gamma=-2Z\, \mathrm{Im}\, \Sigma(E=E_\mathrm{p})$ and $\Omega_\mathrm{R}=\sqrt{Z}\Omega_0$~\cite{kohstall2012metastability}, where $Z\equiv 1\left/\left(1-\frac{\partial \Sigma}{ \partial E}\big|_{E=\Ep}\right)\right.$ is in general interpreted as the quasiparticle residue, which is the overlap between the Fermi polaron wavefunction and the non-interacting impurity wavefunction.

In Fig.~\ref{FIG:3}a, we show $\Gamma$ as a function of the drive strength, along with the $T$-matrix predictions. Focusing at first on the low-drive regime, we do not find a well-pronounced plateau at low $\hbar\Omega_0/\EF$. This is surprising since it has been assumed that for $\hbar\Omega_0/\EF\lesssim 1$, the drive does not affect the polaron~\cite{Scazza2017}. Accessing even lower drive strengths, in order to see the plateau, is experimentally challenging as we approach the overdamped regime $\Omega_\R/\Gamma<0.5$, in which our method for extracting $\Omega_\mathrm{R}$ is unreliable. Note that a previous measurement at a given $\Omega_0$ is in good agreement with our data (red square in Fig.~\ref{FIG:3}~\cite{Scazza2017}). For stronger drives, $\hbar\Omega_0\gtrsim E_\mathrm{F}$, the decay rate shows an unexpected non-monotonic behavior. First, it increases by over an order of magnitude between our lowest $\Omega_0$ and a maximum located around $\hbar\Omega_0/E_\mathrm{F}\approx 3$; it then decreases as a power law of the drive strength. 
 
We use the $T$-matrix approach to calculate $\Gamma$, shown as a black solid line in Fig.~\ref{FIG:3}a. It is in qualitative agreement with the data but is quantitatively imprecise for all but the largest drives. The power law at large drive strengths originates from the properties of the two-body scattering: scattered particles have an energy $\propto \hbar\Omega_0$ and the $T$ matrix is dominated by the two-body contribution, so that it is proportional to the inverse square root of $\Omega_0$~\cite{Sagi2022}. Our data is in good agreement with the corresponding analytical prediction $\hbar\Gamma/\EF\approx 16/\left(3\pi \sqrt{\hbar\Omega_0/\EF}\right)$ (grey line in Fig.~\ref{FIG:3}a). Motivated by this observation, we introduce a two-body $T$-matrix approximation that neglects in-medium scattering effects but still takes into account the presence of the Fermi sea in the self-energy (black dotted line in Fig~\ref{FIG:3}a). Surprisingly, this approximation captures the data remarkably well for all drive strengths. 

The linear-response spectrum provides another measure for the quasiparticle decay rate. From our measurements (green points in Fig.~\ref{FIG:2}c), we extract a width $\Gamma_\text{lin}=0.19(2)  \,\EF/\hbar$, in good agreement with a previous measurement~\cite{yan2019boiling}. This width is significantly larger than our weakest-drive $\Gamma=0.072(5)E_\mathrm{F}/\hbar$ (see Fig.~\ref{FIG:3}a). However, it is important to note that injection/ejection spectra and Rabi oscillations are differently affected by Fourier broadening and finite impurity concentration effects~\cite{Hu2022}. Therefore, their respective relationship to the linear-response properties in the impurity limit is not straightforward and remains to be clarified~\cite{kohstall2012metastability,Adlong2020}. 

The normalized (dynamical) Rabi frequency $\Omega_\text{R}/\Omega_0$ extracted from the damped oscillations also depends on the bare Rabi frequency $\Omega_0$. In Fig.~\ref{FIG:3}b, we show $\Omega_\text{R}/\Omega_0$ versus $\hbar\Omega_0/\EF$. At $\hbar\Omega_0/\EF\approx 2$, $\Omega_\text{R}/\Omega_0$ abruptly increases, reaches a maximum exceeding unity around $\hbar\Omega_0/\EF\approx 5$, and then decreases towards one. In the regime of strong drives, we find $\Omega_\mathrm{R}>\Omega_0$. An explanation for this effect could have been that the minimum of $\OmegaR$ is not reached at the detuning $\Delta=\Delta_0$. 
However, we measure $\Omega_\mathrm{R}$ as a function of the detuning (see inset in Fig.~\ref{FIG:3}b) and find that $\Omega_\mathrm{R}$ is consistently larger than $\Omega_0$; its minimum is reached at $\Delta=1.03(5)\Delta_0$. 

While this observation is at odds with the expectation that $(\Omega_\mathrm{R}/\Omega_0)^2$ is the quasiparticle residue (which is upper-bounded by one), we find that both the two-body and the in-medium $T$-matrix approximations predict that the formally-defined $Z\equiv 1/\left(1-\frac{\partial \Sigma}{ \partial E}\big|_{E=\Ep}\right)$ follows a similar behavior (the dotted and solid black lines in Fig.~\ref{FIG:3}b, respectively). In particular, we analytically calculate $Z\approx 1+16/(6\pi (\hbar\Omega_0/\EF)^{3/2})>1$ in the limit $\hbar\Omega_0/\EF\gg 1$ (see SM~\cite{supp}). This prediction, shown as the grey line in Fig.~\ref{FIG:3}b, is in very good agreement with our measurements. The fact that $Z>1$ indicates a breakdown of the interpretation of the driven impurity interacting with the bath as a well defined quasiparticle. This is in contrast to the usual criterion of $Z\rightarrow 0$ for the disappearance of a quasiparticle, which, for instance, applies to the case for the immobile Fermi impurity~\cite{PhysRevA.84.063632,PhysRevX.2.041020}.

\begin{figure}
  \centering
  \includegraphics[width=1\columnwidth]{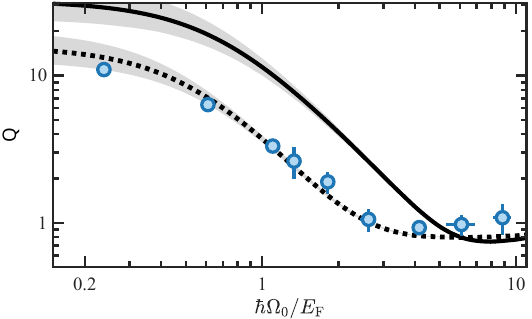}
  \caption{\textbf{Quasiparticle quality factor of the driven polaron.} The quality factor $\Q\equiv\Delta_0/\Gamma$ of the driven polaron measured in the bare-state basis $\{\ket{\uparrow},\ket{\downarrow}\}$ is extracted from the data of Figs.~\ref{FIG:3}a and S8~\cite{supp} (by binning). The solid (dotted) line is the theoretical prediction using the in-medium (two-body) $T$ matrix with error bands indicating the experimental uncertainty in the temperature.  
  The horizontal error bars correspond to the standard deviation of the drive strength in each bin, and the vertical error bars are calculated from the uncertainties of the fits within each bin. 
  } 
  \label{FIG:4}
\end{figure}

This finding motivates us to examine the quasiparticle quality factor $\Q$, defined as the ratio of the quasiparticle energy and its decay rate.
Using the zero crossing as a proxy for the drive-dressed polaron energy, we define $\Q\equiv\Delta_0/\Gamma$. We show $\Q$ versus $\hbar\Omega_0/E_\textrm{F}$ in Fig.~\ref{FIG:4} and find that $\Q\gg 1$ for low $\Omega_0$, indicating a well-defined polaron. However, the quality factor decreases by over an order of magnitude to a minimum of $\approx 0.9$ at $\hbar \Omega_0/\EF\approx 4$, indicating that the polaron is no longer well defined, specifically, it is close to the critical-damping threshold $\Q=1/2$. 

The fact that $\Q\approx 1$ and that $(\Omega_\mathrm{R}/\Omega_0)^2>1$ at $\hbar \Omega_0/\EF\approx 4$ signals that the traditional quasiparticle picture fails in that regime. In general, the quasiparticle residue is defined through a Taylor expansion of the self-energy around the polaron energy $E_\mathrm{p}$. However, the thus-obtained $Z$ yields nonphysical values for the quasiparticle residue. We indeed see that within the $T$-matrix calculations, $Z$ depends on the energy at which it is evaluated, even within the spectral width $\Gamma$; this goes beyond the usual quasiparticle paradigm (see SM~\cite{supp}). 
Nevertheless, it is intriguing that even though $Z$ ($>1$) is no longer the quasiparticle residue in the strong-drive regime, its formal expression is in good agreement with the measured $(\Omega_\textrm{R}/\Omega_0)^2$.

So far, we have probed how the properties of driven polarons formed in the bare-state basis ($\ket{\uparrow}$ and $\ket{\downarrow}$) change with drive strength. However, in the large-drive limit, it is more natural to consider the driven polarons formed by atoms in the (rf-)dressed states $\frac{1}{\sqrt{2}}\left(\ket{\uparrow}\pm \ket{\downarrow}\right)$ interacting with the Fermi gas. The energy of these \emph{dressed-state polarons} is approximately given by $E_\mathrm{p}\pm \hbar \Omega_{0}/2$ while their decay rate is $\propto 1/\sqrt{\hbar \Omega_0/\EF}$ for $\hbar\Omega_0/E_\mathrm{F}\gg1$ (see SM~\cite{supp}), leading to a large quality factor at large drives. However, we found from our theoretical analysis 
that the spectral functions of the dressed-state polarons around $\hbar\Omega_0/\EF\approx 4$ (where $\Q$ is smallest) significantly differ from a Lorentzian quasiparticle line shape, showing that the quasiparticle breakdown is also reflected in the dressed-state polaron. In that regime, the dressed-state polaron hybridizes with a continuum -- the remnant at the unitary limit of the repulsive polaron -- which suggests that the quasiparticle breakdown observed here is due to the drive-induced merging of a coherent excitation with an incoherent background (see SM~\cite{supp}). 

In conclusion, we have shown that the driven impurity in a Fermi gas is a powerful platform for studying non-equilibrium quantum dynamics. In the future, it could be used to study many-body systems without well-defined quasiparticles, such as the spin-balanced unitary Fermi gas in the normal phase~\cite{PhysRevLett.122.203402}, or the heavy impurity system~\cite{PhysRevLett.111.265302,PhysRevA.104.043309}. It would also be interesting to directly observe the dressed-state polarons with a secondary probing field. Furthermore, in the regime $1/\kF a_{\uparrow\B}>0$, one could observe richer steady-state magnetization spectra for pre-thermal states containing both attractive and repulsive polarons~\cite{PhysRevLett.111.265302}. Theoretically, it is imperative to elucidate the connection between the impurity's equilibrium spectral properties and Rabi oscillations, \emph{e.g.} by using a kinetic approach~\cite{wasak2022decoherence} or a fully self-consistent $T$-matrix calculation.

%%%%%%%%%%%%%%%%%%%%%%
%%  Acknowledgements %%
%%%%%%%%%%%%%%%%%%%%%%
We thank Ehud Altman, Fr\'{e}d\'{e}ric Chevy, Eleanor Crane, Eugene Demler, Zoran Hadzibabic, Paul Julienne, Clemens Kuhlenkamp, Jesper Levinsen, Brendan Mulkerin, Meera Parish, and Richard Schmidt for fruitful discussions. We thank Francesco Scazza and Giacomo Roati for sharing their experimental data. This work was supported by the NSF (Grant Nos. PHY-1945324 and PHY-2110303), DARPA (Grant No. W911NF2010090), the David and Lucile Packard Foundation, and the Alfred P. Sloan Foundation. G.L.S acknowledges support from the NSF Graduate Research Fellowship Program. A.S. acknowledges support from the U.S. Department of Energy, Office of Science, National Quantum Information Science Research Centers, Quantum Systems Accelerator and the DOE Office of Science, Office of Advanced Scientific Computing Research (ASCR) Quantum Computing Application Teams program, under fieldwork proposal number ERKJ347.
M.K. acknowledges support from the Deutsche Forschungsgemeinschaft (DFG, German Research Foundation) under Germany's Excellence Strategy--EXC--2111--390814868, DFG grants No. KN1254/1-2, KN1254/2-1, and TRR 360 - 492547816 and from the European Research Council (ERC) under the European Unions Horizon 2020 research and innovation programme (Grant Agreement No. 851161), as well as the Munich Quantum Valley, which is supported by the Bavarian state government with funds from the Hightech Agenda Bayern Plus.
\vspace{-1.em}

\begingroup
\renewcommand{\addcontentsline}[3]{}% Remove functionality of \addcontentsline
\section{Methods}
\textbf{Preparation of the highly-imbalanced uniform Fermi gas.} We prepare an incoherent mixture of the first and third lowest Zeeman sub-levels (denoted $\ket{\uparrow}$, $\ket{\B}$) of $^6$Li atoms in a red-detuned optical dipole trap. The internal states of the impurity species are $\ket{\uparrow}\equiv\ket{\frac{1}{2},+\frac{1}{2}}$, $\ket{\downarrow}\equiv\ket{\frac{1}{2},-\frac{1}{2}}$ and the bath internal state is $\ket{\B}\equiv\ket{\frac{3}{2},-\frac{3}{2}}$, in the $\ket{F,m_F}$ basis at low magnetic field ($F$ and $m_F$ are the total spin and its projection along the magnetic field axis). We evaporatively cool this mixture at a magnetic field $B\approx \unit[284]{G}$, where the s-wave scattering length between $\ket{\uparrow}$ and $\ket{\B}$ is $a_{\uparrow \B}\approx-900 a_0$. The atoms in state $\ket{\uparrow}$ are then transferred into $\ket{\downarrow}$ with a \unit[5]{ms} Landau-Zener rf sweep. After a \unit[100]{ms} hold, we adiabatically ramp the field to $B\approx\unit[583]{G}$ in $\unit[500]{ms}$, where $a_{\downarrow \B}\approx 0$, and then load the atoms into a blue-detuned optical box trap, formed by the intersection of two beams (of wavelength $\unit[639]{nm}$) shaped by digital micromirror devices. The radius and length of the cylindrical box are $R=\unit[63(1)]{\mu m}$ and $L = \unit[121(1)]{\mu m}$, respectively. We adjust the concentration $x\equiv N_{\downarrow}/N_\mathrm{B}$ (where $N_\B$ is the bath atom number) to be typically $0.17(2)$, by optically blasting atoms in $\ket{\downarrow}$ with
a $\unit[12]{\mu s}$ light pulse of controllable intensity. Finally, we ramp the field to $B_0\approx\unit[690]{G}$ and hold it for $\unit[400]{ms}$ for equilibration. Typically, we have $N_\B\approx 5\times10^5$ atoms in the bath state, and the bath temperature is $T= 0.25(2)T_\mathrm{F}$ (measured by time-of-flight expansion with the weakly interacting mixture $\ket{\downarrow}$-$\ket{\B}$). To estimate the uncertainty due to the interaction between the impurity and the bath, we ramp the field back to \unit[583]{G} to measure the temperature, which is within the error bar of the one extracted at \unit[690]{G}.

\textbf{Extracting the renormalized Rabi frequency $\Omega_{\R}$ and decay rate $\Gamma$.}
To extract $\Omega_{\R}$ and $\Gamma$ on resonance, we fit the time evolution of the magnetization with $\mathcal{M}(t)=-\cos(\Omega_{\R}t)\exp(- \Gamma t/2)$. 
The off-resonant Rabi oscillation in Fig.~\ref{FIG:1}b is fitted with a phenomenological model $\mathcal{M}(t)=\mathcal{M}_\infty(\tilde{\Delta})-A\exp(-\Gamma t/2)\cos(\OmegaR t)-(1-A+\mathcal{M}_\infty(\tilde{\Delta}))\exp(-\Gamma^\prime t/2)$, where $\mathcal{M}_\infty(\tilde{\Delta})$ is the asymptotic magnetization and $\Gamma$ and $\Gamma^\prime$ are decay rates.

 \textbf{Model Hamiltonian of the system.} 
  As the largest detuning used in the experiment ($\lesssim\unit[100]{kHz}$) is much smaller than the transition frequency between the $\ket{\uparrow}$ and $\ket{\downarrow}$ states ($\approx\unit[76.0]{MHz}$), we can use the rotating-wave approximation. Within this approximation, the energy is conserved in the frame rotating with the drive. The Hamiltonian in that frame is
 \begin{equation}
 \begin{aligned}
 & H=  \frac{\hbar\Omega_0}{2} \sum_{\mathbf{k}}( c_{\mathbf{k}\downarrow}^\dagger  c_{\mathbf{k}\uparrow} + \mathrm{h.c.})+ \sum_{\mathbf{k}}  (\epsilon_\mathbf{k}+\hbar\Delta)c^\dagger_{\mathbf{k}\downarrow}  c_{\mathbf{k}\downarrow}+\sum_{\mathbf{k}}   \epsilon_\mathbf{k}   c^\dagger_{\mathbf{k}\uparrow}  c_{\mathbf{k}\uparrow}\notag \\
 &+\sum_{\mathbf{k}}(\epsilon_\mathbf{k}-\mu)   d^\dagger_{\mathbf{k}}  d_{\mathbf{k}} +\frac{1}{\mathcal{V}} \sum_{\alpha=\uparrow,\downarrow} g_\alpha\sum_{\mathbf{k},\mathbf{k'},\mathbf{q}}   c^\dagger_{\mathbf{k'}+\mathbf{q}\alpha}  c_{\mathbf{k'}\alpha} d^\dagger_{\mathbf{k}-\mathbf{q}}  d_{\mathbf{k}},\notag
 \end{aligned}
 \end{equation}
 where $\epsilon_\mathbf{k}=\hbar^2\mathbf{k}^2/(2m)$, $m$ is the atom's mass, $\mu$ is the chemical potential of the $\ket{\mathrm{B}}$ atoms and $\mathcal{V}$ is the volume of the system; $c^\dag_{\mathbf{k}\alpha}$ ($c_{\mathbf{k}\alpha}$) is the creation (annihilation) operator of state $\alpha\in\lbrace \uparrow, \downarrow\rbrace$ with momentum $\hbar\mathbf{k}$; $d^\dag_{\mathbf{k}}$ and $d_{\mathbf{k}}$ are the creation and annihilation operators for a particle in the bath $\ket{\B}$ with momentum $\hbar\mathbf{k}$. The coupling constants $g_\alpha$ are connected to the s-wave scattering lengths $a_{\alpha \mathrm{B}}$ by the Lippmann-Schwinger equation in the infinite volume limit, see SM~\cite{supp}. In the single-impurity limit, we can neglect interactions between atoms in the states $\ket{\uparrow}$ and $\ket{\downarrow}$. 

\textbf{Extraction of the zero crossing $\Delta_0$.} 
The zero crossing is extracted without \emph{a priori} knowledge of the functional form of $\mathcal{M}$ using a linear fit in the range $\mathcal{M}\in[-0.3,0.3]$ in the vicinity of $\mathcal{M} = 0$. Additionally, the data is fitted with the $T>0$ extension of model Eq.~\eqref{eq:M}, using the bath temperature $T$ $(\approx T_\text{spin})$ as an input parameter. These two methods have been checked to be consistent. We perform at least three repetitions for each data point for the steady-state spectrum around the zero crossing.

\textbf{Zero crossing from spectral functions.} We calculate the magnetization using the exact relation $N_{\alpha} = \int \mathrm{d}\omega (\sum_\mathbf{q} A_{\alpha\alpha}(\mathbf{q},\omega))/(e^{\beta(\hbar\omega-\mu_\mathrm{imp})}+1)$ between the populations $N_\alpha$ (where $\alpha\in\lbrace \uparrow, \downarrow\rbrace$) and the equilibrium spectral functions $A_{\alpha\alpha}(\mathbf{q},\omega)$ in the presence of the drive; $\mu_\mathrm{imp}$ is the impurity chemical potential. In the single-impurity limit $\mu_\mathrm{imp}\rightarrow-\infty$, and it thus drops out of the expression for $\mathcal{M}$. The resulting expression is evaluated using the equilibrium spectral functions from the $T$-matrix approximation. The zero crossing is then extracted as in the experiment, from a linear fit to $\mathcal{M}(\Delta)$. To get a qualitative understanding of the $T$-matrix results, we compare them to the zero crossings obtained using a generic quasiparticle ansatz for the spectral functions~\cite{Adlong2020,supp}. In this case, we find that $\mathcal{M} =\left.\left((\Delta-\frac{E_\mathrm{p}}{\hbar}) \tanh\left(\frac{\beta \OmegaR}{2}\right)+\frac{Z-1}{Z+1}\OmegaR\right)\middle/\right.\left(\OmegaR+\frac{Z-1}{Z+1}(\Delta-\frac{E_\mathrm{p}}{\hbar})\tanh\left(\frac{\beta \OmegaR}{2}\right)\right)$ with $\Omega_\R\approx\sqrt{Z\Omega_0^2+(\Delta-\frac{E_\mathrm{p}}{\hbar})^2}$. This reduces to the $T>0$ generalization of Eq.~\eqref{eq:M} with $Z=1$. We obtain an analytical expression for the zero crossing in the weak- and strong-drive limits:
\begin{equation}
        \hbar\Delta_0=
        \begin{cases}
        E_\mathrm{p} - \frac{1}{2}\hbar\Omega_0(Z-1)&\hbar\Omega_0\gg E_\F\\
        E_\mathrm{p} + 2\kB T \mathrm{arctanh}\left(\frac{1-Z}{Z+1}\right)&\hbar\Omega_0\ll E_\F.
    \end{cases}\notag
\end{equation}
In particular, for $\hbar\Omega_0\ll E_\F$ we find $\hbar\Delta_0=E_\mathrm{p}$ at $T=0$. However, away from this limit, the zero crossing is shifted away from the polaron energy.

\textbf{Effect of nonzero $\kF a_{\downarrow \B}$.} In our $T$-matrix approximation, we take into account the effect of a nonzero $\kF a_{\downarrow \B}$. In the experiment, $\kF a_{\downarrow \B}$ is small and positive, leading to the existence of a repulsive polaron in the $\ket{\downarrow}$ state. The dominant effect on the zero crossing $\Delta_0$ is a shift of the polaron energy by the ($\ket{\downarrow}$ state) repulsive polaron energy~\cite{yan2019boiling} (see also SM~\cite{supp}). The effect of nonzero $\kF a_{\downarrow \B}$ on the decay rate $\Gamma$ and the renormalized Rabi frequency $\Omega_\R$ shown in Fig.~\ref{FIG:3} is visible at large $\hbar\Omega_0/E_\F$ limit, where we set $\kF a_{\downarrow \B}=0$ for the calculation of the exact limit. Thus, the discrepancy between the black lines and the grey line in Fig.~\ref{FIG:3} is due to the nonzero $\kF a_{\downarrow \B}$ (note that the temperature effect can be neglected in the limit $\hbar\Omega_0/E_\F\gg1$). 

%%%%%%%%%%%%%%%%%%%%%%
%%  BIBLIOGRAPHY    %%
%%%%%%%%%%%%%%%%%%%%%%

\endgroup

\newpage
\clearpage
\definecolor{blue(pigment)}{rgb}{0.2, 0.2, 0.6}
\definecolor{darkerblue}{rgb}{0.0, 0.0, 0.4}
\definecolor{darkblue}{rgb}{0.0,0.0,0.5}
\definecolor{darkgreen}{rgb}{0.0,0.4,0.0}
\hypersetup{
   colorlinks,
   linkcolor=blue(pigment),
    citecolor=darkgreen,
   urlcolor=darkblue
}

\setcounter{figure}{0}
\renewcommand{\thefigure}{S\arabic{figure}}%
\setcounter{equation}{0}
\renewcommand{\theequation}{S\arabic{equation}}
\setcounter{secnumdepth}{3}

\onecolumngrid
% \title{Supplementary Material \\The strongly driven Fermi polaron}
% \normalsize	
{\centering\bf  \large Supplementary material for:\\
\vspace{0.1cm}
\centering
The strongly driven Fermi polaron\\}
% \maketitle
\onecolumngrid
\tableofcontents
\vspace{1em}

\section{$T$-matrix theory for the strongly driven Fermi polaron}

In this section, we define the model Hamiltonian and derive the $T$-matrix theory presented in the main text. We then calculate the impurity properties in the presence of the drive using the $T$-matrix formalism. We interpret the properties of the resulting driven Fermi polaron using a quasiparticle ansatz. Finally, we show how to connect the $T$-matrix predictions with the magnetization measured in the experiment.

\subsection{Model Hamiltonian of a driven impurity immersed in a Fermi gas}

We consider the limit of a single impurity with two internal states, interacting with a Fermi gas (the bath) via contact interactions. In this section alone, we work in the natural units $\EF=\kF=k_\B=\hbar=1$, where $\kF,\, \EF$ are the Fermi wavevector and energy of the bath, $\kB$ is the Boltzmann constant, and $\hbar$ is the reduced Plank's constant. We label the impurity states as $\ket{\uparrow}$ and $\ket{\downarrow}$ and the bath state as $\ket{\mathrm{B}}$. A radio-frequency (rf) drive of Rabi frequency $\Omega_0$ is detuned by $\Delta$ with respect to the bare transition frequency between the states $\ket{\uparrow}$ and $\ket{\downarrow}$ (\emph{i.e.} in the absence of $\ket{\B}$). The s-wave scattering length between states $\ket{\uparrow}$ and $\ket{\mathrm{B}}$ diverges, $1/a_{\uparrow \mathrm{B}}= 0$. Following the typical experimental parameters, we use $a_{\downarrow \mathrm{B}}= 0.16$ for the scattering length between states $\ket{\downarrow}$ and $\ket{\mathrm{B}}$, and a temperature $\beta^{-1}=0.3$ for the bath (unless otherwise indicated). Even for the largest detunings used in the experiment, $\Delta$ is much smaller than the transition frequency between $\ket{\uparrow}$ and $\ket{\downarrow}$, such that we can use the rotating-wave approximation. In the rotating frame of the drive, the (time-independent) Hamiltonian of the system is 
\begin{equation}
H=  \frac{\Omega_0}{2} \sum_{\mathbf{k}}( c_{\mathbf{k}\downarrow}^\dagger  c_{\mathbf{k}\uparrow} + \mathrm{h.c.})+ \sum_{\mathbf{k}} (\epsilon_\mathbf{k}+\Delta) c^\dagger_{\mathbf{k}\downarrow}  c_{\mathbf{k}\downarrow}+\sum_{\mathbf{k}}   \epsilon_\mathbf{k}   c^\dagger_{\mathbf{k}\uparrow}  c_{\mathbf{k}\uparrow} 
+\sum_{\mathbf{k}}   (\epsilon_\mathbf{k}-\mu)  d^\dagger_{\mathbf{k}}  d_{\mathbf{k}} +\frac{1}{\mathcal{V}} \sum_{\alpha=\uparrow,\downarrow} g_\alpha\sum_{\mathbf{k},\mathbf{k'},\mathbf{q}}   c^\dagger_{\mathbf{k'}+\mathbf{q}\alpha}  c_{\mathbf{k'}\alpha} d^\dagger_{\mathbf{k}-\mathbf{q}}  d_{\mathbf{k}},
\label{eq:Hamiltonian}
\end{equation}
where $\epsilon_\mathbf{k}=\mathbf{k}^2/2m$ is the kinetic energy, $\mu$ is the chemical potential of the $\ket{\mathrm{B}}$ atoms and $\mathcal{V}$ is the volume of the system. The operators $c^\dag_{\mathbf{k}\alpha}$ and $c_{\mathbf{k}\alpha}$ are the creation and annihilation operators for the impurity in state $\alpha \in \lbrace \uparrow, \downarrow \rbrace$ with momentum $\mathbf{k}$; $d^\dag_{\mathbf{k}}$ and $d_{\mathbf{k}}$ are the creation and annihilation operators for a particle in the bath $\ket{\B}$ with momentum $\mathbf{k}$.

The coupling constants $g_\alpha$ are related to the scattering lengths $a_{\alpha \mathrm{B}}$ by the Lippmann-Schwinger equation in the infinite volume limit:
\begin{equation}
    \frac{1}{g_\alpha}=\frac{1}{8\pi a_{\alpha \mathrm{B}}}-\int^{\Lambda}\frac{\mathrm{d}^3 \mathbf{k}}{(2\pi )^3} \frac{1}{2\mathbf{k}^2},
    \label{eq:LippmannSchwinger}
\end{equation}
where $\Lambda$ is a momentum cutoff. Below, we explicitly eliminate the cutoff and evaluate all expressions in the limit $\mathcal{V} \rightarrow \infty$ and $\Lambda\rightarrow \infty$. 

The Rabi coupling $\Omega_0$ mixes the states $\ket{\uparrow}$ and $\ket{\downarrow}$ such that they do not form the natural basis for this problem, especially in the limit of large $\Omega_0$. 
Instead, we use the states dressed by the drive to diagonalize the Hamiltonian in the absence of interactions with the bath. The transformation to the dressed-state operators $ c_{\mathbf{k}\pm}$ can be written as $( c_{\mathbf{k}+},  c_{\mathbf{k}-})^T=V ( c_{\mathbf{k}\uparrow},  c_{\mathbf{k}\downarrow})^T$, where the transformation matrix is
\begin{equation}
V= \frac{1}{\sqrt{2\Omega_{\mathrm{R},0}}}
	\begin{pmatrix}
		\sqrt{\Omega_{\mathrm{R},0}-\Delta} & \sqrt{\Omega_{\mathrm{R},0}+\Delta}\\
		\sqrt{\Omega_{\mathrm{R},0}+\Delta} & -\sqrt{\Omega_{\mathrm{R},0}-\Delta}
	\end{pmatrix},
\end{equation}
and the effective Rabi frequency is $\Omega_{\mathrm{R},0}=\sqrt{\Omega_0^2+\Delta^2}$. Note that $V^\dagger=V^{-1}=V$. In the dressed-state basis, the Hamiltonian becomes
\begin{equation}
 H=  \frac{\Omega_{\mathrm{R},0}}{2} \sum_{\mathbf{k}}( c_{\mathbf{k}+}^\dagger  c_{\mathbf{k}+} -  c_{\mathbf{k}-}^\dagger  c_{\mathbf{k}-}) +\sum_{\mathbf{k},\sigma= +,-}   \left(\epsilon_\mathbf{k}+\frac{\Delta}{2}\right)   c^\dagger_{\mathbf{k}\sigma}  c_{\mathbf{k}\sigma}+\sum_{\mathbf{k}}   (\epsilon_\mathbf{k}-\mu)  d^\dagger_{\mathbf{k}}  d_{\mathbf{k}}+\frac{1}{\mathcal{V}} \sum_{\mathbf{k},\mathbf{k'},\mathbf{q}}  
 ( c^\dagger_{\mathbf{k'}+\mathbf{q}+}  c^\dagger_{\mathbf{k'}+\mathbf{q}-})\tilde U \begin{pmatrix} 
 c_{\mathbf{k'+}}\\
 c_{\mathbf{k'-}}
\end{pmatrix} d^\dagger_{\mathbf{k}-\mathbf{q}}  d_{\mathbf{k}}, \label{eq:dressed_H}
\end{equation}
where the interaction matrix $\tilde U$ in the dressed-state basis is
\begin{equation}
	\tilde U= \frac{1}{2}
	\begin{pmatrix}
	g_\uparrow+g_\downarrow-\frac{\Delta}{\Omega_{\mathrm{R},0}}\left(g_\uparrow-g_\downarrow \right) & \frac{\Omega_0}{\Omega_{\mathrm{R},0}}\left(g_\uparrow-g_\downarrow \right)\\
	\frac{\Omega_0}{\Omega_{\mathrm{R},0}}\left(g_\uparrow-g_\downarrow \right) & g_\uparrow+g_\downarrow+\frac{\Delta}{\Omega_{\mathrm{R},0}}\left(g_\uparrow-g_\downarrow \right)
	\end{pmatrix}.
 \label{eq:intmatrix}
\end{equation}
By changing to the dressed-state basis, we diagonalize the quadratic part of the Hamiltonian with respect to the internal states (at the expense of making the quartic part non-diagonal).

\subsection{Impurity Green's function}

As we will show below, the experimentally relevant observables for this work can be computed from the impurity Green's function:
	\begin{equation}
	\hat G^{-1}(\mathbf{k},\omega)=\begin{pmatrix}
	\omega-\epsilon_\mathbf{k} &-\Omega_0/2\\
	-\Omega_0/2 & \omega-\epsilon_\mathbf{k}-\Delta
	\end{pmatrix}-\hat \Sigma(\mathbf{k},\omega)
	\label{eq:GF},
\end{equation}
where we use a hat to denote a matrix-valued quantity in the bare-state basis.

To determine the self-energy matrix $\hat \Sigma$, we use the non-self-consistent $T$-matrix approximation: 
\begin{equation}
	\hat \Sigma(\mbf{k},\omega) \approx\int \frac{\mathrm{d}^3 \mbf{q}}{(2\pi)^3} \hat T(\mbf{k}+\mbf{q},\omega+(\epsilon_\mbf{q}-\mu)) n(\mbf{q})
	\label{eq:selfenergy},
\end{equation}
where $n(\mbf{q})=\frac{1}{e^{\beta(\mathbf{q}^2-\mu)}+1}$ is the Fermi-Dirac distribution for the bath. This approximation describes well the ground state properties of the polaron in the absence of the drive~\cite{PhysRevLett.98.180402S}.

\subsection{$T$ matrix}
Inspired by the related two-dimensional problem tackled in Ref.~\cite{Kuhlenkamp2022S} (where $a_{\downarrow\B}= 0$), we calculate the $T$ matrix for the three-dimensional problem in a similar fashion, taking into account that $ a_{\downarrow\B}\neq 0$ (see also Ref.~\cite{Hu2022S} for a related calculation). We first write the $T$ matrix in the dressed-state basis
\begin{equation}
	\tilde T(\mathbf{k},\omega)^{-1}=\tilde U^{-1}-\tilde \Pi(\mathbf{k},\omega)
	\label{eq:tildeT}
\end{equation}
in terms of the non-self-consistent pair propagator matrix $\tilde{\Pi}$ (also in the dressed-state basis), whose elements $\tilde \Pi_{\sigma\sigma'}$ (with $\sigma,\sigma'\in\lbrace +,-\rbrace$) are:
\begin{equation}
	\tilde \Pi_{\sigma\sigma'}(\mathbf{k},\omega)=\int\frac{\mathrm{d}^3 \mbf{q}}{(2\pi)^3} \frac{\delta_{\sigma\sigma'} (1-n(\mathbf{q}))}{\omega-E_\sigma+\mu-\epsilon_\mathbf{q} -\epsilon_{\mbf{q}-\mbf{k}}+i\eta},
\end{equation}
where $\eta\rightarrow0^+$ and \begin{equation}
E_{\pm}=\frac{1}{2}(\Delta\pm \Omega_{\mathrm{R},0}).
\end{equation}

Inserting the Lippmann-Schwinger equation Eq.~\eqref{eq:LippmannSchwinger} for $g_\sigma$ into $\tilde U$ in Eq.~\eqref{eq:intmatrix}, we remove the cutoff dependence in the pair propagator. We then calculate $\tilde T$ by inverting the right-hand side of Eq.~\eqref{eq:tildeT}. Finally, by transforming the $T$ matrix back into the bare-state basis ($\{\ket{\uparrow},\ket{\downarrow}\}$), $\hat T\equiv V \tilde T V$, we get
\begin{equation}\label{eq:Tmatrix}
	\frac{\hat T  (\omega)}{8\pi a_{\downarrow \mathrm{B}}T_0 (\omega)}=
\begin{pmatrix}

		\frac{1}{4\pi a_{\downarrow \mathrm{B}}}- \left(1-\frac{\Delta}{\Omega_{\mathrm{R},0}}\right)\Pi(\omega-E_-)-\left(1+\frac{\Delta}{\Omega_{\mathrm{R},0}}\right)\Pi(\omega-E_+)  &  \frac{\Omega_0}{\Omega_{\mathrm{R},0}}\left[\Pi(\omega-E_+)-\Pi(\omega-E_-)\right] \\
		\frac{\Omega_0}{\Omega_{\mathrm{R},0}}\left[\Pi(\omega-E_+)-\Pi(\omega-E_-)\right] & 
 
  \frac{1}{4\pi a_{\uparrow \mathrm{B}}}-\left(1+\frac{\Delta}{\Omega_{\mathrm{R},0}}\right)\Pi(\omega-E_-)-\left(1-\frac{\Delta}{\Omega_{\mathrm{R},0}}\right)\Pi(\omega-E_+) 

	\end{pmatrix}
	,
\end{equation}
where
\begin{equation}
\begin{aligned}
	\frac{1}{T_0 (\omega)}=\frac{1}{4\pi a_{\uparrow \mathrm{B}}}-\left(1+\frac{a_{\downarrow \mathrm{B}}}{a_{\uparrow \mathrm{B}}}\right)\left(\Pi(\omega-E_+)+\Pi(\omega-E_-) \right)&+\frac{\Delta}{\Omega_{\mathrm{R},0}}\left(1-\frac{a_{\downarrow \mathrm{B}}}{a_{\uparrow \mathrm{B}}}\right)\left(\Pi(\omega-E_+)-\Pi(\omega-E_-)\right)\\
     &+16\pi a_{\downarrow \mathrm{B}} \Pi(\omega-E_+)\Pi(\omega-E_-).
 \end{aligned}
\end{equation}
We dropped the momentum in the arguments for brevity (since all $\omega$-dependent quantities are evaluated at the same momentum) and expressed the $T$ matrix in terms of the explicitly renormalized pair propagator
	\begin{align}	\Pi(\omega)\equiv\Pi(\mathbf{k},\omega)&=\int\frac{\mathrm{d}^3 \mbf{q}}{(2\pi)^3}\frac{1-n(\mathbf{q})}{\omega-(\epsilon_\mathbf{q}-\mu) -\epsilon_{\mbf{q}-\mbf{k}}+i\eta}-\int\frac{\mathrm{d}^3 \mbf{q}}{(2\pi)^3} \frac{1}{2\mbf{q}^2} \\
	&=-\frac{i\sqrt{\omega-k^2/2+\mu}}{8\pi\sqrt{2}}-\int \frac{\mathrm{d}^3 \mbf{q}}{(2\pi)^3}\frac{n(\mathbf{q})}{\omega-(\epsilon_\mathbf{q}-\mu) -\epsilon_{\mbf{q}-\mbf{k}}+i\eta}.
	\label{eq:Pi}
\end{align}
In the last step, we inserted the exact solution of the two-body pair propagator~\cite{Hu2022_2S}. 

We thus find that it is the pair propagator \emph{without the drive} $\Pi$ that enters the $T$ matrix in Eq.~\eqref{eq:Tmatrix}; it is simply evaluated at energies shifted by $E_\pm$. The remaining integral in Eq.~\eqref{eq:Pi} represents the in-medium part of the pair propagator and can be solved semi-analytically even at finite temperature~\cite{Hu2022_2S}. In the following, we will distinguish the in-medium $T$ matrix from the two-body $T$ matrix. The former is evaluated using both terms in the pair propagator while the latter only contains the first term of Eq.~\eqref{eq:Pi}.

To gain some insight into the rather involved $T$-matrix expression in Eq.~\eqref{eq:Tmatrix}, we now discuss three limits.

\subsubsection{Limit of small Rabi frequency: $\Omega_0\rightarrow 0$} 

In the limit $\Omega_0\rightarrow 0 $ and $\Delta\rightarrow 0$, the $T$ matrix is diagonal: 
\begin{equation}
	\hat T(\mathbf{k},\omega)^{-1}\stackrel{\Omega_0, \Delta\rightarrow 0}{=} \begin{pmatrix}
		\frac{1}{8\pi a_{\uparrow \mathrm{B}}} -\Pi(\mathbf{k},\omega) & 0 \\
		0 & \frac{1}{8\pi a_{\downarrow \mathrm{B}}} -\Pi(\mathbf{k},\omega)
	\end{pmatrix}.
	\end{equation}
The scattering properties of each state are thus separately renormalized by the scattering with the bath; there is thus no mixing between the two states.

\subsubsection{Limit of vanishing $\ket{\downarrow}$-$\ket{\B}$ interactions: $a_{\downarrow \mathrm{B}}\rightarrow 0$} 
In this case, only the $\uparrow \uparrow$ component of $\hat{T}$ is non-vanishing:
\begin{equation}
	\hat T_{\uparrow\uparrow}(\mathbf{k},\omega)=\frac{1}{\frac{1}{8\pi a_{\uparrow \mathrm{B}}}-\frac{1}{2}\left((1+\frac{\Delta}{\Omega_{\mathrm{R},0}})\Pi(\mathbf{k},\omega-E_-)+(1-\frac{\Delta}{\Omega_{\mathrm{R},0}})\Pi(\mathbf{k},\omega-E_+)\right)}.
	\label{eq:T_zeroa23}
\end{equation}
This implies that only the $\uparrow\uparrow$ component is modified by the interactions with $\ket{\mathrm{B}}$, despite the mixing of the states $\ket{\uparrow}$ and $\ket{\downarrow}$ induced by the drive. The qualitative form of the $T$ matrix is similar to the case without the drive; the pair propagator is simply replaced by a sum of two pair propagators that are evaluated at energies $\omega-E_{\pm}$. In particular, at large $\Omega_0$, the frequency arguments of the pair propagators are dominated by $E_\pm$, implying that in this regime the energy of the scattered particles is given by $\Omega_0$. This has important consequences in the large $\Omega_0$ limit.

\subsubsection{Limit of large Rabi frequency: $\Omega_0\gg 1$\label{ch:largeomega}}

\begin{figure}
	\includegraphics{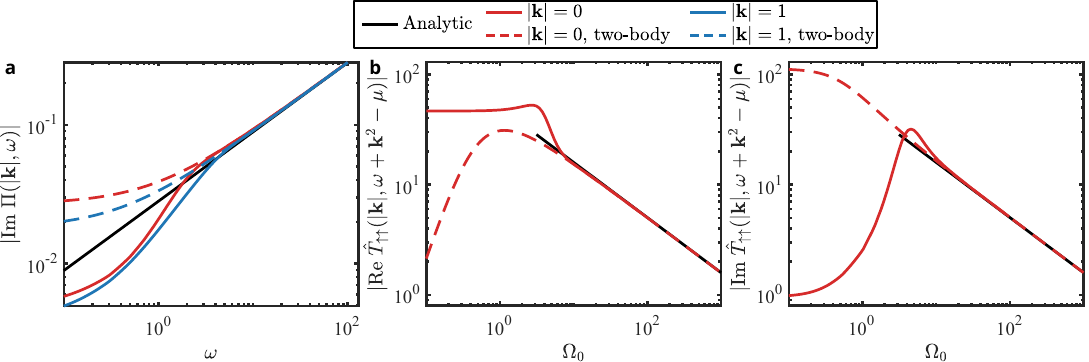}
	\caption{\textbf{Pair propagator and $T$ matrix in the limit of large $\Omega_0$.} (\textbf{a}) Imaginary part of the pair propagator evaluated with both terms in Eq.~\eqref{eq:Pi} (solid colored lines) and with only the two-body part (dashed lines) with momentum $|\mathbf{k}|=0$ and $|\mathbf{k}|=1$ (see legends). We compare the imaginary part of the pair propagator with the analytical large-$\omega$ limit in Eq.~\eqref{eq:Pilargew} (solid black line). (\textbf{b}) Real and (\textbf{c}) imaginary parts of the $\uparrow\uparrow$ component of the $T$ matrix in Eq.~\eqref{eq:T_zeroa23} with the full pair propagator (solid red line) and with the two-body part (dashed red line) for $\omega=0.5,\, |\mathbf{k}|=0$. We evaluated the $T$ matrix at a generic non-zero detuning $\Delta=0.5$, with other detunings showing a similar behavior. We compare the $T$ matrix with the analytical limit in Eq.~\eqref{eq:T_largeom_zeroa23} (solid black line). We set $a_{\downarrow \mathrm{B}}=0$.}\label{Fig:S1}
\end{figure}

 In the limit of large frequencies $\omega\gg 1$, the pair propagator in Eq.~\eqref{eq:Pi} is dominated by the two-body part since the in-medium part is suppressed as $1/\omega$ (see colored lines in Fig.~\ref{Fig:S1}a). In this limit, the pair propagator is~\cite{Sagi2022S}

\begin{equation}
	\Pi(\mathbf{k},\omega) \stackrel{\omega\gg 1,|\mathbf{k}|\approx 1}\approx \frac{-i}{8\pi\sqrt{2}}\sqrt{\omega},
	\label{eq:Pilargew}
\end{equation} 
see the black line in Fig.~\ref{Fig:S1}a. Since the $T$ matrix in Eq.~\eqref{eq:Tmatrix} contains $\Pi(\pm\Omega_0/2)$, both its real and imaginary parts are dominated by the power-law behavior at large $\Omega_0$ (see Fig.~\ref{Fig:S1}b and c). The fact that the interaction $a_{\downarrow \mathrm{B}}$ is nonzero (but weak) leads to two interesting regimes where power laws appear. These two limits are distinguished by the relative strength of $\Omega_0$ with respect to $a_{\downarrow \mathrm{B}}$ and are universal, as neither scattering lengths $a_{\uparrow \mathrm{B}}$ and $a_{\downarrow \mathrm{B}}$ influence the dynamics in these regimes.

In the first limit 
\begin{equation}
\text{Max}( 1/a_{\uparrow\B},1)\ll\sqrt{\Omega_0}/2 \ll  1/a_{\downarrow \B},  \label{eq:largeOmegalimit}  
\end{equation} 
we find that only $\hat T_{\uparrow\uparrow}$ is non-vanishing, and
\begin{equation}
	\hat T_{\uparrow\uparrow}(\mathbf{k},\omega)\approx -\frac{16\pi}{\sqrt{\Omega_0}} (1+i).
	\label{eq:T_largeom_zeroa23}
\end{equation}
This is the regime realized in our experiment, as $1/a_{\uparrow\B}=0$ and $1/a_{\downarrow \B}\approx 6.25 > \sqrt{\Omega_0}/2$ even for our largest  $\Omega_0$ ($\approx 10$). We show this asymptote as the black line in Fig.~\ref{Fig:S1}b and Fig.~\ref{Fig:S1}c and find excellent agreement at large $\Omega_0$ when setting $ a_{\downarrow \mathrm{B}}=0$.

In the second limit (of even larger Rabi frequencies)
\begin{equation}
\sqrt{\Omega_0}/2 \gg  1/a_{\downarrow \B}\label{eq:superlargeOmegalimit},  
\end{equation} 
we find 
\begin{equation}\label{eqn: T matrix at large Omega0 limit}
	\hat T(\mathbf{k},\omega)\approx -\frac{8\pi}{\sqrt{\Omega_0}} \begin{pmatrix} 1+i & 1-i \\
	1-i & 1+i
	\end{pmatrix},
\end{equation}
such that both components scatter off the bath component $\ket{\mathrm{B}}$ with the same strength. 
Note that despite the fact that the magnitude of the off-diagonal terms of the $T$ matrix is the same as the diagonal ones, they only lead to small corrections to the Green's function (see the $\Omega_0/2$ off-diagonal terms in Eq.~\eqref{eq:GF}).

\subsection{Quasiparticle ansatz and spectral functions}

We evaluate the spectral functions in the bare-state basis, defined as
\begin{equation}
    \hat{A}(\mathbf{k},\omega)\equiv-\frac{1}{\pi} \mathrm{Im}\left(\hat G(\mathbf{k},\omega)\right). \label{eq:specfunc}
\end{equation}
These spectral functions contain information about the excitations of the system in the presence of the drive. They can be directly measured by probing the steady state in the presence of the drive with a second rf/microwave pulse (that would couple the dressed impurity to another -- empty -- internal state). In our experiment, the Rabi oscillations provide an indirect probe of these spectral functions, see Sec.~\ref{sct:Rabifromspec}.

\subsubsection{Quasiparticle properties}\label{section: Qp properties}
In order to interpret the spectral functions obtained from the $T$ matrix, we use a quasiparticle ansatz~\cite{Adlong2020S}. We define the quasiparticle ansatz for the Green's function as
\begin{equation}
	\left(\hat G^{\mathrm{QP}}\right)^{-1}(\mathbf{k},\omega)=\begin{pmatrix}
	\frac{1}{Z_\uparrow}\left(\omega-\left(\frac{m}{m^*_\uparrow}\epsilon_{\mathbf{k}}+E_{\mathrm{p}\uparrow}\right)+i\frac{\Gamma_\uparrow}{2}\right) &-\frac{\Omega_0}{2}\\
	-\frac{\Omega_0}{2} & \frac{1}{Z_\downarrow}\left(\omega-\left(\frac{m}{m^*_\downarrow}\epsilon_{\mathbf{k}}+\Delta+E_{\mathrm{p}\downarrow}\right)+i\frac{\Gamma_\downarrow}{2}\right)
	\end{pmatrix},
	\label{eq:GFqp}
\end{equation}
where the ($\Omega_0$- and $\Delta$-dependent) quasiparticle properties are defined as
\begin{align}
	E_{\mathrm{p}\alpha}&=\mathrm{Re}(\hat \Sigma_{\alpha\alpha}(\mathbf{k}=0,E_{\mathrm{p}\alpha})),\notag\\
	Z_\alpha&= \frac{1}{1-\frac{\partial \mathrm{Re}(\hat\Sigma_{\alpha\alpha}(\mathbf{k}=0,\omega)}{\partial \omega}\bigg|_{\omega=E_{\mathrm{p}\alpha}}},\notag\\
	\Gamma_\alpha&=-2Z_\alpha\,\mathrm{Im}(\hat\Sigma_{\alpha\alpha}(\mathbf{k}=0,\omega=E_{\mathrm{p}\alpha})),\notag\\
	\frac{m^*_\alpha}{m} &= \frac{Z_\alpha}{1+\frac{\partial \mathrm{Re}(\hat\Sigma_{\alpha\alpha}(\mathbf{k}=0,\omega))}{\partial \epsilon_\mathbf{p}}\bigg|_{\omega=E_{\mathrm{p}\alpha}}},\label{eq:QP_props}
\end{align}
where $\hat \Sigma_{\alpha\alpha}$ is the $\alpha\alpha$ component of the self-energy matrix $\hat \Sigma$. As usual for a quasiparticle ansatz approach, the momentum dependence of the self-energy is neglected. Moreover, we assume that the off-diagonal component $\hat \Sigma_{\uparrow\downarrow}(\mathbf{k}=0,\omega)$ is negligible. Indeed, we numerically find that $|\hat \Sigma_{\uparrow\downarrow}(\mathbf{k}=0,\omega)|/\Omega_0\leq 4\%$ for $\Omega_0\leq 100$ and $a_{\downarrow \mathrm{B}}=0.16$, justifying this approximation. 

\begin{figure}
	\includegraphics[width=\columnwidth]{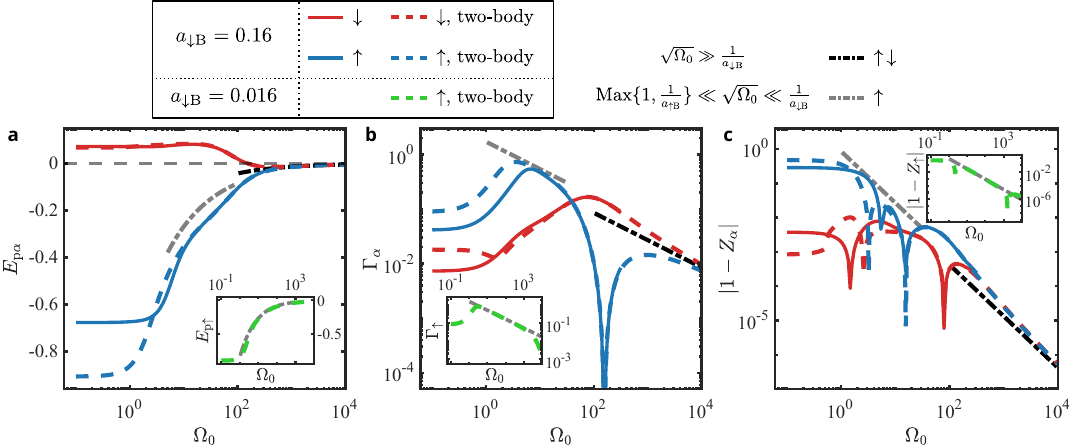}
	\caption{\textbf{Driven Fermi polaron properties.} (\textbf{a}) Drive-dressed polaron energy $E_{\mathrm{p}\alpha}$. (\textbf{b}) Decay rate $\Gamma_\alpha$. (\textbf{c}) Quasiparticle residue $Z_\alpha$ ($\alpha\in\{\uparrow,\downarrow\}$), where $Z_\alpha<1$ for small $\Omega_0$ and each resonance-like feature in the plot corresponds to a sign change of $1-Z_\alpha$. In the main figures, we set $a_{\downarrow \mathrm{B}}=0.16$ and compare results using the in-medium (solid lines) or only two body (dashed lines) $T$-matrix expressions. The black (gray) dash-dotted lines are the exact results evaluated in two limits (see legends). In the insets, we show the comparison between two-body results (green dashed lines) for $a_{\downarrow \mathrm{B}}=0.016$ and the large $\Omega_0$ limit (gray dash-dotted lines).
    Because of the connection to the Rabi oscillation properties of $\Gamma$ and $Z$, (see Sec.~\ref{sct:Rabifromspec}), the prediction for $\ket{\uparrow}$ (blue lines) corresponds to the one we show in the main text. The $E_\mathrm{p}$ shown in the main text and below corresponds to $E_{\mathrm{p}\uparrow}-E_{\mathrm{p}\downarrow}$.\label{Fig: 2S}}
\end{figure}

We calculate the quasiparticle properties from Eq.~\eqref{eq:QP_props} using the drive-dependent self-energy Eq.~\eqref{eq:selfenergy} and fixing $\Delta=E_{\mathrm{p}\uparrow}-E_{\mathrm{p}\downarrow}$ (see Fig.~\ref{Fig: 2S}). While we pick $\Delta=\Delta_0$ in the experiment, the former choice is numerically more convenient; we checked that the results are essentially indistinguishable. The self-energy is either evaluated using the two-body (dashed lines) or the in-medium (solid lines) $T$ matrix.

We also compare the numerical results with the analytical large-$\Omega_0$ limit. We again distinguish the two regimes discussed in section~\ref{ch:largeomega}. We show the first regime,  $\mathrm{Max}\left(1/a_{\uparrow \mathrm{B}},1\right)\ll \sqrt{\Omega_0}/2 \ll  1/a_{\downarrow \mathrm{B}}$, as gray dash-dotted lines. Inserting Eq.~\eqref{eq:T_largeom_zeroa23} into the self-energy Eq.~\eqref{eq:selfenergy}, we find 
\begin{equation}
	E_{\mathrm{p}\uparrow}=-\frac{8}{3\pi\sqrt{\Omega_0}}.
 \label{eq:largeOmEp}
\end{equation}

To calculate the quasiparticle residue, we analytically evaluate the derivative of the self-energy Eq.~\eqref{eq:selfenergy} and take the limit of large $\Omega_0$. We find
\begin{equation}
	Z_\uparrow\approx 1+\frac{8}{3\pi}\Omega_0^{-\frac{3}{2}}.
  \label{eq:largeOmZ}
\end{equation}
This shows that the quasiparticle residue approaches one from \emph{above}. Finally, using the fact that $\mathrm{Im}(\hat\Sigma_{\uparrow\uparrow})=\mathrm{Re}(\hat\Sigma_{\uparrow\uparrow})$ in this limit, we get
\begin{equation}
	\Gamma_\uparrow\approx -2E_{\mathrm{p}\uparrow}. 
\end{equation}

In this limit, the analytical expressions (shown as the gray dash-dotted lines in Fig.~\ref{Fig: 2S} and gray solid lines in Fig.~3 of the main text) do not agree very well with the $T$-matrix results, because of the effect of nonzero $a_{\downarrow \mathrm{B}}$. Indeed, using a smaller $a_{\downarrow \mathrm{B}}$ ($a_{\downarrow \mathrm{B}}=0.016$ for the green dashed lines in the insets of Fig.~\ref{Fig: 2S}), the agreement is greatly improved.

In the second regime, $\sqrt{\Omega_0} \gg  1/a_{\downarrow \mathrm{B}}$ (shown as black dash-dotted lines in Fig.~\ref{Fig: 2S}), both components are equal in the $T$ matrix Eq.~\eqref{eqn: T matrix at large Omega0 limit}. We find
 \begin{align}
     E_{\mathrm{p}\uparrow} &=  E_{\mathrm{p}\downarrow} = -\frac{4}{3\pi\sqrt{\Omega_0}},\\
     Z_\uparrow &=  Z_\downarrow = 1-\frac{4}{3\pi\Omega_0^{3/2}}\label{eq: Z large W0 limit},\\
     \Gamma_\uparrow &=  \Gamma_\downarrow = \frac{8}{3\pi\sqrt{\Omega_0}}.
 \end{align}
 
We find good agreement between the analytical and numerical results for $\Omega_0\gtrsim 10^3$. In this regime, all quasiparticle properties obey power laws with $\Omega_0$ that are universal, in the sense that they do not depend on the temperature and $a_{\downarrow \mathrm{B}}$ (the only other energy scales in this system apart from $\EF$). 

We find that $Z_\uparrow$ crosses one twice. Indeed $Z_\uparrow<1$ for small $\Omega_0$. In the intermediate regime where $\mathrm{Max}\left(1/a_{\uparrow \mathrm{B}},1\right)\ll \sqrt{\Omega_0}/2 \ll  1/a_{\downarrow \mathrm{B}}$ (the one realized in the experiment), $Z_\uparrow>1$, see Eq.~\eqref{eq:largeOmZ}. Finally $Z_\uparrow<1$ for $\sqrt{\Omega_0}\gg1/a_{\downarrow\B}$, see Eq.~\eqref{eq: Z large W0 limit}. 

Moreover, ${\Gamma_\uparrow}$ drops to zero at $\Omega_0=\Omega_0^\text{c}\approx150$ (see blue lines in Fig.~\ref{Fig: 2S}b). Inserting Eq.~\eqref{eq:Pilargew} into Eq.~\eqref{eq:QP_props}, we analytically find $\Omega_0^\text{c}=4/a_{\downarrow \mathrm{B}}^2\approx156$. This prediction agrees very well with the numerical evaluation. This oddity arises due to the interplay between the drive strength and the nonzero $a_{\downarrow \mathrm{B}}$. The relationship $\Omega_0^\text{c}\propto 1/a_{\downarrow \mathrm{B}}^2$ is reminiscent of the expression of a Feshbach molecule's energy in terms of the scattering length, but the origin of this phenomenon remains to be understood. 

\begin{figure}
	\includegraphics[width=0.5\columnwidth]{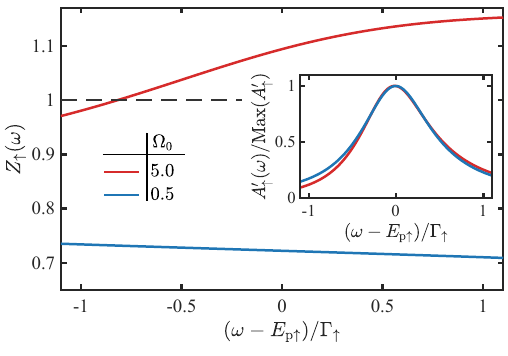}
	\caption{\textbf{Frequency-dependent $Z_\uparrow$.} $Z_\uparrow(\omega)$ as defined in Eq.~\eqref{eq:QP_props} is shown versus $\omega$ (here, we set $a_{\downarrow\B}=0$ and $\Delta=E_{\mathrm{p}\uparrow}$). Inset: Decoupled spectral functions $A'_\uparrow(\omega)$ for the values of $\Omega_0$ shown in the main panel. To emphasize the minute differences, $A'_\uparrow$ are normalized to their respective peak values. For reference $\text{Max}(A'_\uparrow)=8.9$ (resp. $0.93$) and $\Gamma_\uparrow=0.05$ (resp. $0.74$) for $\Omega_0=0.5$ (resp. $5$).
 \label{Fig:S3}}
\end{figure}

To explain the unusual result that $Z_\uparrow>1$, we examine the assumptions underlying the quasiparticle ansatz approach (Eqs.~\eqref{eq:GFqp}-\eqref{eq:QP_props}). In particular, the quasiparticle ansatz assumes that $\partial\mathrm{Re}(\hat\Sigma_{\uparrow\uparrow}(\mathbf{k}=0,\omega)/\partial\omega$ is independent of the frequency over the width of the quasiparticle peak. To test this assumption, we formally define a frequency-dependent residue $Z_\uparrow(\omega)\equiv 1/\left( 1-\frac{\partial \mathrm{Re}(\hat\Sigma_{\uparrow\uparrow}(\mathbf{k}=0,\omega)}{\partial \omega}\right)$ as well as a decoupled spectral function $A'_\uparrow(\omega)\equiv-\frac{1}{\pi}\, \mathrm{Im}\left(1/\left(\hat{G}^{-1}\right)_{\uparrow\uparrow}\right)$, and display them in Fig.~\ref{Fig:S3}. We find that for $\Omega_0=0.5$, $Z_\uparrow(\omega)$ is indeed roughly constant over the width of the quasiparticle peak. By contrast, for $\Omega_0=5$, $Z_\uparrow(\omega)$ significantly varies and exceeds one at large $\omega$. The traditional assumption of a frequency-independent quasiparticle residue is thus not valid (and the spectral function is not Lorentzian shaped, see inset of Fig.~\ref{Fig:S3}). As a result, the $Z$ defined in Eq.~\eqref{eq:QP_props} (which is not necessarily constrained to $<1$) differs from the spectral weight, \emph{i.e.} the integral of the spectral function over the quasiparticle peak. 

\subsubsection{Quasiparticle spectral functions}

We now calculate the quasiparticle-ansatz spectral functions $
\hat A^{\mathrm{QP}}(\mathbf{k},\omega)\equiv-\frac{1}{\pi} \mathrm{Im}(\hat G^{\mathrm{QP}}(\mathbf{k},\omega))$ from Eq.~\eqref{eq:GFqp}. We first focus on the zero-momentum part, and the calculation for the various components of $\hat A^{\mathrm{QP}}$ yields: 
 \begin{align}
\hat A^{\mathrm{QP}}_{\uparrow\uparrow}(\omega)=\frac{Z}{2\pi|\OmegaR|^2} 
    &\Bigg [ \left( \left(\frac{\Gamma}{2} \Omega^\mathrm{r}_R -\OmegaR^\mathrm{i}\tilde \Delta \right)\frac{\omega-\omega_+^\mathrm{r}}{\omega_+^\mathrm{i}}+\tilde \Delta \OmegaR^\mathrm{r}-|\OmegaR|^2+\OmegaR^\mathrm{i} \frac{\Gamma}{2}\right)\frac{\omega_+^\mathrm{i}}{(\omega-\omega_+^\mathrm{r})^2+(\omega_+^\mathrm{i})^2}\notag\\
    &- \left( \left(\frac{\Gamma}{2} \Omega^\mathrm{r}_R -\OmegaR^\mathrm{i}\tilde \Delta \right)\frac{\omega-\omega_-^\mathrm{r}}{\omega_-^\mathrm{i}}+\tilde \Delta \OmegaR^\mathrm{r}+|\OmegaR|^2+\OmegaR^\mathrm{i} \frac{\Gamma}{2}\right)\frac{\omega_-^\mathrm{i}}{(\omega-\omega_-^\mathrm{r})^2+(\omega_-^\mathrm{i})^2}
    \Bigg],\label{eq:specfuncupup}\\
    \hat A^{\mathrm{QP}}_{\downarrow\downarrow}(\omega)=-\frac{1}{2\pi|\OmegaR|^2} &\Bigg [ \left( \left(\frac{\Gamma}{2} \Omega^\mathrm{r}_R -\OmegaR^\mathrm{i}\tilde \Delta \right)\frac{\omega-\omega_+^\mathrm{r}}{\omega_+^\mathrm{i}}+\tilde \Delta \OmegaR^\mathrm{r}+|\OmegaR|^2+\OmegaR^\mathrm{i} \frac{\Gamma}{2}\right)\frac{\omega_+^\mathrm{i}}{(\omega-\omega_+^\mathrm{r})^2+(\omega_+^\mathrm{i})^2}\notag\\
    &- \left( \left(\frac{\Gamma}{2} \Omega^\mathrm{r}_R -\OmegaR^\mathrm{i}\tilde \Delta \right)\frac{\omega-\omega_-^\mathrm{r}}{\omega_-^\mathrm{i}}+\tilde \Delta \OmegaR^\mathrm{r}-|\OmegaR|^2+\OmegaR^\mathrm{i} \frac{\Gamma}{2}\right)\frac{\omega_-^\mathrm{i}}{(\omega-\omega_-^\mathrm{r})^2+(\omega_-^\mathrm{i})^2}
    \Bigg],
    \label{eq:specfunc_22}\\
    \hat A^{\mathrm{QP}}_{\uparrow\downarrow}(\omega)=\frac{Z\Omega_0}{2\pi|\OmegaR|^2} &\left[\left(\OmegaR^\mathrm{i}\frac{\omega-\omega_+^\mathrm{r}}{\omega_+^\mathrm{i}}-\OmegaR^\mathrm{r}\right)\frac{\omega_+^\mathrm{i}}{(\omega-\omega_+^\mathrm{r})^2+(\omega_+^\mathrm{i})^2}-\left(\OmegaR^\mathrm{i}\frac{\omega-\omega_-^\mathrm{r}}{\omega_-^\mathrm{i}}-\OmegaR^\mathrm{r}\right)\frac{\omega_-^\mathrm{i}}{(\omega-\omega_-^\mathrm{r})^2+(\omega_-^\mathrm{i})^2}\right]\label{eq:specfunc_12},
\end{align}
where the superscripts `r' and `i' denote the real and imaginary parts, and
\begin{align}
E_\mathrm{p}&=E_{\mathrm{p}\uparrow}-E_{\mathrm{p}\downarrow},\\
\tilde \Delta&=\Delta-E_\mathrm{p},\\
    \omega_\pm &= E_{\mathrm{p}\uparrow}+\frac{1}{2}\left(\tilde \Delta\pm\Omega^\mathrm{r}_\mathrm{R}+i\left(\pm\Omega^\mathrm{i}_\mathrm{R}-\frac{\Gamma}{2}\right)\right),\label{eq: pole position}\\
    \OmegaR &=e^{i\theta/2}\left((Z\Omega_0^2-(\Gamma/2)^2+\tilde \Delta^2)^2+(\Gamma\tilde \Delta)^2\right)^{1/4},\\
    \theta &= \arctan\left(\frac{\Gamma\tilde \Delta}{Z\Omega_0^2-(\Gamma/2)^2+\tilde \Delta^2} \right).\label{eq:theta}
\end{align}

For clarity, we have set in Eqs.~\eqref{eq:specfuncupup}-\eqref{eq:theta} -- and for the rest of the supplementary material -- $\Gamma\equiv\Gamma_\uparrow$, $Z\equiv Z_\uparrow $, and assumed that $\Gamma_\downarrow=0$, and $Z_\downarrow=1$. The corresponding expressions for $\mathbf{k}\neq 0$ can be obtained with the substitution $E_{\mathrm{p}\alpha}\rightarrow E_{\mathrm{p}\alpha} +(m/m^*_\alpha) \epsilon_\mathbf{k}$. 

\begin{figure}
    \centering
    \includegraphics[width=\columnwidth]{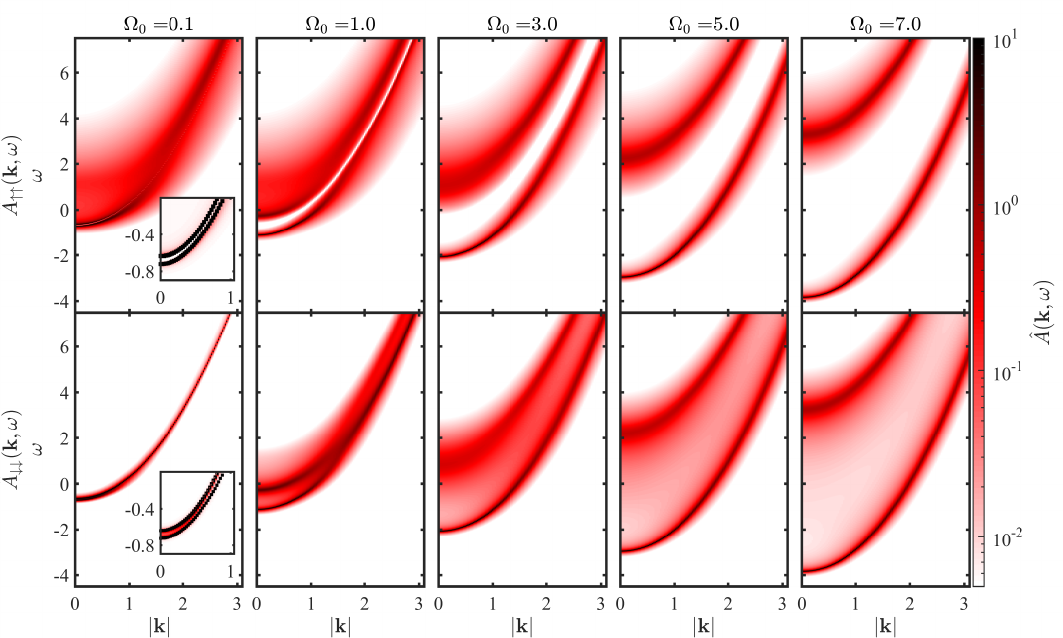}
    \caption{\textbf{Spectral functions $\hat A_{\uparrow\uparrow}(\mathbf{k},\omega)$ (upper row) and $\hat A_{\downarrow\downarrow}(\mathbf{k},\omega)$ (lower row) for different Rabi frequencies.} Insets in the first column ($\Omega_0=0.1$) are the same plots with a linear color scale. The black points in the insets are the quasiparticle dispersion relation Eq.~\eqref{eq:om_momdep}. The color scale is the magnitude of the spectral function. In this figure, we set $a_{\downarrow \mathrm{B}}=0$,  $\Delta=E_{\mathrm{p}\uparrow}$, and $E_{\mathrm{p}\downarrow}=0$.\label{Fig:S4}}
    
\end{figure}

\begin{figure}
    \centering
    \includegraphics[width=\columnwidth]{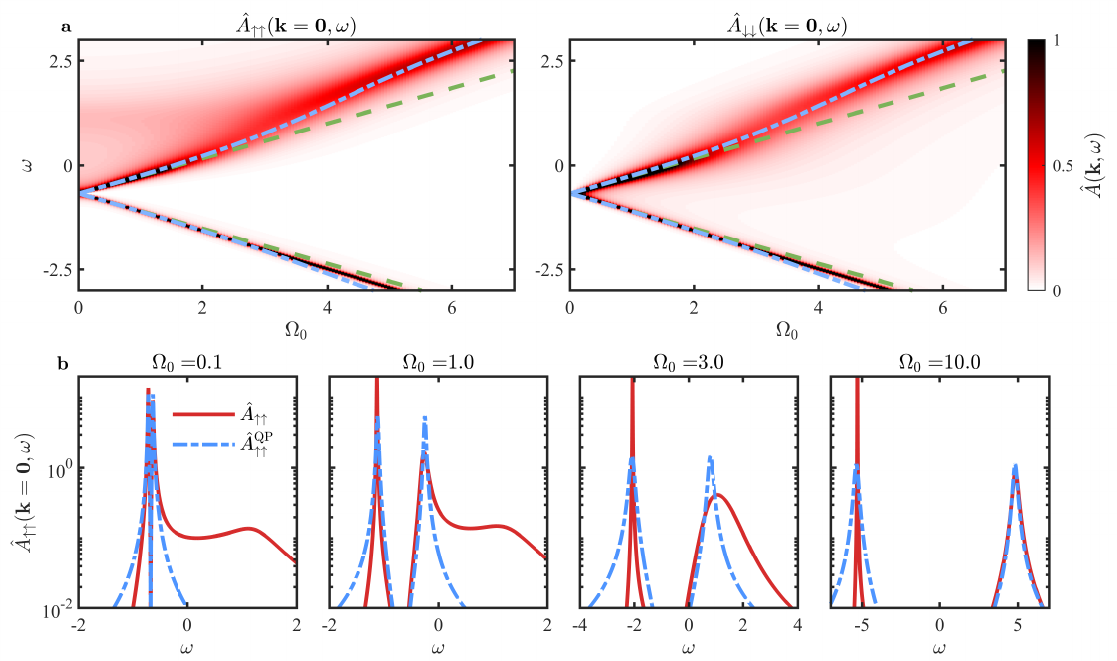}
    \caption{\textbf{Zero-momentum spectral functions.} (\textbf{a}) $\hat A_{\uparrow\uparrow}(\mathbf{k}=0,\omega)$ and $\hat A_{\downarrow\downarrow}(\mathbf{k}=0,\omega)$. The lines are the quasiparticle dispersions $\omega_{\pm}^\mathrm{r}(\mathbf{k}=\mathbf{0})=E_{\mathrm{p}\uparrow}\pm\frac{1}{2}\sqrt{Z}\Omega_0$, where we neglect (green dashed lines) or take into account (blue dash-dotted lines) the $\Omega_0$ dependence of $Z$ and $E_\mathrm{p}$. The color scale is the magnitude of the spectral function. (\textbf{b}) $\hat A_{\uparrow\uparrow}(\mathbf{k}=0,\omega)$ (red solid lines) and $\hat A^{\mathrm{QP}}_{\uparrow\uparrow}(\mathbf{k}=0,\omega)$ (blue dash-dotted lines), evaluated at different $\Omega_0$. In this figure, we set $a_{\downarrow \mathrm{B}}=0$, $\Delta=E_{\mathrm{p}\uparrow}$, and $E_{\mathrm{p}\downarrow}=0$.\label{Fig:S5}}
    
\end{figure}

\subsubsection{$T$-matrix spectral function vs. quasiparticle spectral function}\label{compsrison of spectral functions}

We compare predictions using the quasiparticle ansatz to those directly derived from the $T$ matrix. In Fig.~\ref{Fig:S4}, we show the $T$-matrix spectral functions $\hat A_{\uparrow\uparrow}(\mathbf{k},\omega)$ and $\hat A_{\downarrow\downarrow}(\mathbf{k},\omega)$ for different $\Omega_0$. We find two sharp peaks with a nearly quadratic dispersion relation, signaling a spectral function that contains two quasiparticles, the \emph{dressed-state polarons}. Indeed, setting $\tilde\Delta=0$ in Eq.~\eqref{eq: pole position}, and assuming that $\Gamma\ll\Omega_0$, we expect the position of the peaks in the quasiparticle spectral functions to be:

\begin{equation}
    \omega_\pm^\mathrm{r}(\mathbf{k})=E_{\mathrm{p}\uparrow}\pm\frac{1}{2}\sqrt{Z\Omega_0^2 +\mathbf{k}^4\left(1-\frac{m}{m^*}\right)^2} + \mathbf{k}^2\frac{1}{2}\left(1+\frac{m}{m^*}\right),
    \label{eq:om_momdep}
\end{equation}  
where we assumed $m^*_\downarrow=m$ and defined $m^*\equiv m^*_\uparrow $. 

We test Eq.~\eqref{eq:om_momdep} at small $\Omega_0$ and small momenta by comparing the peak positions of the $T$-matrix spectral functions with Eq.~\eqref{eq:om_momdep} using the effective mass of the attractive polaron  ($m^*/m=1.18$~\cite{Scazza2017S}), see the black points in the insets of Fig.~\ref{Fig:S4} (where $\Omega_0=0.1$). We find excellent agreement with the peaks of the spectral functions shown as the red color scale.

Next, we examine $\omega_\pm^\mathrm{r}(\mathbf{k}=\mathbf{0})$ with respect to $\Omega_0$. In Fig.~\ref{Fig:S5}a, we plot the zero-momentum spectral functions $\hat A_{\uparrow\uparrow}(\mathbf{k}=\mathbf{0},\omega)$ and $\hat A_{\downarrow\downarrow}(\mathbf{k}=\mathbf{0},\omega)$. To test Eq.~\eqref{eq:om_momdep} in that sector, we first set $E_{\mathrm{p}\uparrow}$ and $Z$ to their values at $\Omega_0=0$ (green dashed lines), finding a good agreement for small $\Omega_0$. However, for $\Omega_0\approx 3$, the upper branch shifts upwards when the incoherent background merges into it. This background, which is the `remnant' at unitarity of the repulsive polaron (a well-defined quasiparticle for weaker interactions) is only visible in $\hat A_{\uparrow\uparrow}$. 

If we instead use in $\omega_\pm^\mathrm{r}$ the values of $E_\mathrm{p}(\Omega_0)$ and $Z(\Omega_0)$ determined from Eq.~\eqref{eq:QP_props}  (blue dash-dotted lines), the agreement is improved. This shows that the peak positions are nonlinear in $\Omega_0$, due to the dependence of $E_\mathrm{p}$ and $Z$ on $\Omega_0$.

\begin{figure}
    \centering
    \includegraphics[width=\columnwidth]{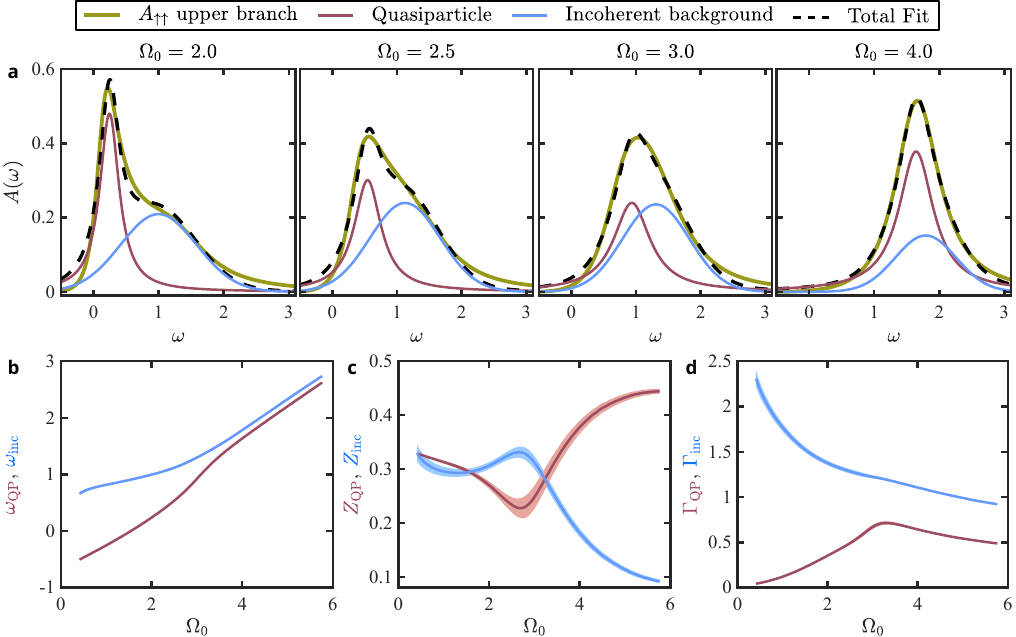}
    \caption{\textbf{Merger of the upper-branch dressed-state polaron with the incoherent background.} (\textbf{a}) Upper branch of $\hat A_{\uparrow\uparrow}(\mathbf{k}=0,\omega)$ for $\Omega_0=2,\,2.5,\,3,$ and $4$. The yellow solid lines are the spectral functions obtained from the $T$ matrix (Eq.~\eqref{eq:specfunc}). The red (blue) solid lines are fits to the quasiparticle peak (incoherent background), modelled as a Lorentzian (a Gaussian). The black dashed lines are the sum of the fitted spectral functions. (\textbf{b}) Peak positions  ($\omega_\text{QP}$ and $\omega_\text{inc}$), (\textbf{c}) spectral weight ($Z_\text{QP}$, and $Z_\text{inc}$), and (\textbf{d}) full width at half maximum  ($\Gamma_\text{QP}$ and $\Gamma_\text{inc}\equiv2\sqrt{2\ln2}\sigma_\text{inc}$) as a function of $\Omega_0$ for the quasiparticle and the incoherent parts respectively, see Eq.~\eqref{eq:sumQPinc}. The colored bands correspond to standard errors of the fit. \label{Fig:S6}}
    
\end{figure}

While the quasiparticle ansatz agrees qualitatively with the $T$-matrix approach, we find quantitative discrepancies. We examine these differences by plotting the $T$-matrix (red solid lines) and quasiparticle (blue dash-dotted lines) spectral functions at $\mathbf{k}=0$ in Fig.~\ref{Fig:S5}b for various $\Omega_0$. We find good agreement for the peak positions for all $\Omega_0$. However, the large $|\omega|$ wings of the two spectral functions are qualitatively different. Moreover, while the peak heights of the two dressed states are different in the $T$-matrix spectral functions, they are the same for the quasiparticle ansatz. This can be directly seen in the expressions for the quasiparticle spectral functions in Eq.~(\ref{eq:specfuncupup},~\ref{eq:specfunc_22}): for $\tilde \Delta=0$, and $\Gamma\ll\Omega_0$, we find
\begin{equation}
    \hat{A}^{\mathrm{QP}}_{\uparrow\uparrow}(\omega)=Z \hat A^{\mathrm{QP}}_{\downarrow\downarrow}(\omega)\approx \frac{Z}{8\pi} \left(\frac{\Gamma}{(\omega-\omega_+^\mathrm{r})^2+(\Gamma/4)^2}+\frac{\Gamma}{(\omega-\omega_-^\mathrm{r})^2+(\Gamma/4)^2} \right).
\end{equation}
This result is clearly at odds with the $T$-matrix one, which displays strongly non-Lorentzian line shapes in the upper branch (see the plot for $\Omega_0=3$ in Fig.~\ref{Fig:S5}b). For even larger $\Omega_0$, the upper branch $T$ matrix calculation again agrees well with the quasiparticle spectral function while this is not true for the lower branch (see the plot for $\Omega_0=10$ in Fig.~\ref{Fig:S5}b).

To examine further the merger of the upper-branch dressed-state polaron with the incoherent background, we model the upper branch of the spectral function by the sum of a quasiparticle ansatz and an incoherent (Gaussian) background 
\begin{equation}
    A(\omega)=\frac{Z_\mathrm{QP}}{\pi}\frac{\Gamma_\mathrm{QP}/2}{(\omega-\omega_\mathrm{QP})^2+(\Gamma_\mathrm{QP}/2)^2}+\frac{Z_\mathrm{inc}}{\sqrt{2\pi}\sigma_\mathrm{inc}}\exp{\left(-\frac{(\omega-\omega_\mathrm{inc})^2}{2\sigma_\mathrm{inc}^2}\right)},\label{eq:sumQPinc}
\end{equation}
where $\omega_\mathrm{QP}$ ($\omega_\mathrm{inc}$) is the peak position, $Z_\mathrm{QP}$ ($Z_\mathrm{inc}$) is the spectral weight, and $\Gamma_\mathrm{QP}$ ($2\sqrt{2\ln2}\sigma_\text{inc}$) is the full width at half maximum of the quasiparticle (incoherent background).
This model fits the upper branch of the spectrum $\hat{A}_{\uparrow\uparrow}$ reasonably well for $1<\Omega_0<6$ (see Fig.~\ref{Fig:S6}a). We find that as $\Omega_0$ increases, the quasiparticle peak shifts towards the incoherent background. Simultaneously, the spectral weight of the quasiparticle peak decreases. At $\Omega_0\approx3$ (see Fig.~\ref{Fig:S6}b) the quasiparticle peak has fully merged with the incoherent background. When further increasing $\Omega_0$, the spectra become sharper and the quasiparticle contribution again dominates. A small incoherent background remains for $\Omega_0>4$, which is also reflected in the slightly asymmetric line shape of the spectral function. This simple model provides evidence that the sudden change of quasiparticle properties in this region is due to the merging of a coherent excitation (the dressed-state polaron) with an incoherent background (the remnant of the repulsive polaron). This phenomenon could be further explored by studying the spectral functions away from unitarity ($1/a_{\uparrow \mathrm{B}}\neq 0$), where the repulsive polaron is a well-defined quasiparticle for $\Omega_0=0$.

These findings indicate that the effect of the drive is not merely a renormalization of the quasiparticle properties. In particular, the modification of the line shapes is important in the evaluation of the steady-state magnetization, which we discuss next.

\subsection{Steady-state magnetization}

In this section, we calculate the steady-state magnetization of the driven Fermi polaron. We first start with the simple model of a dressed spin $1/2$ in thermal equilibrium with a bath. We next express the many-body steady-state magnetization in terms of the impurity spectral function. We numerically evaluate the magnetization and the zero crossing $\Delta_0$ using both the $T$-matrix and quasiparticle spectral functions. We compare the numerical zero crossing to the experimental measurements presented in the main text.

\subsubsection{Single-particle magnetization: a dressed spin $1/2$ in thermal equilibrium}

Consider a spin $1/2$ of energy $\omega_0$ dressed by an oscillating field of frequency $\omega$ and Rabi frequency $\Omega_0$. In the frame co-rotating with the field (and with an appropriate choice of phase), the Hamiltonian is

\begin{equation}
    H=-\frac{\Delta}{2}\sigma_z+\frac{\Omega_0}{2}\sigma_x ,
\end{equation}
where $\Delta=\omega_0-\omega$, and $\sigma_x$ and $\sigma_z$ are the Pauli operators. When the dressed spin $1/2$ is in thermal equilibrium with a bath at the temperature $\beta^{-1}$, the partition function is $\Z=\Tr\left[e^{-\beta H}\right]=2\mathrm{cosh}\left(\beta \Omega_{\mathrm{R},0}/2\right)$ in terms of the effective Rabi frequency $\Omega_{\mathrm{R},0}=\sqrt{\Omega_0^2+\Delta^2}$. The spin's magnetization $\mathcal{M}\equiv\frac{1}{\Z}\Tr\left[ e^{-\beta H} \sigma_z\right]$ is then 
\begin{equation}
       \mathcal{M}=\frac{2}{\beta \mathcal{Z}} \frac{d \mathcal{Z}}{d \Delta}
        =\frac{\Delta}{\Omega_{\mathrm{R},0}}\tanh\left(\frac{\beta \Omega_{\mathrm{R},0}}{2}\right). \label{eq:mag_pheno}
\end{equation}

In the following, we show how this simple model emerges from a first-principle treatment of the impurity problem.

\subsubsection{Many-body magnetization}\label{MB steady-state}
The magnetization of the impurity is defined as
\begin{equation}
    \mathcal{M}=\frac{n_\uparrow-n_\downarrow}{n_\uparrow+n_\downarrow},
\end{equation}
where $n_\uparrow$ and $n_\downarrow$ are the densities of the $\ket{\uparrow}$ and $\ket{\downarrow}$ component, respectively (so that $n_\uparrow+n_\downarrow$ is the total impurity density). We consider the single-impurity limit, in which both $n_\uparrow$ and $n_\downarrow$ vanish as the impurity chemical potential $\mu_\mathrm{imp}\rightarrow -\infty$, while the magnetization remains finite.

Using the fluctuation-dissipation relation, the densities of the $\ket{\uparrow}$ and $\ket{\downarrow}$ components are expressed in terms of the spectral functions as  
\begin{align}
    n_{\alpha} &= \frac{1}{\mathcal{V}}\sum_\mathbf{q}\int \mathrm{d}\omega \frac{ \hat A_{\alpha\alpha}(\mathbf{q},\omega)}{e^{\beta(\omega-\mu_\mathrm{imp})}+1} \\
    &\stackrel{\mu_\mathrm{imp}\rightarrow -\infty}{\rightarrow} e^{\beta\mu_\mathrm{imp}}\frac{1}{\mathcal{V}}\sum_\mathbf{q}\int  e^{-\beta\omega} \hat A_{\alpha\alpha}(\mathbf{q},\omega)\mathrm{d}\omega,
\end{align}
where $\alpha \in \lbrace \uparrow, \downarrow \rbrace$. Hence, the magnetization of the impurity is 
\begin{equation}
    \mathcal{M}=\frac{\sum_\mathbf{q}\int  e^{-\beta\omega} \left(\hat A_{\uparrow\uparrow}(\mathbf{q},\omega)-\hat A_{\downarrow\downarrow}(\mathbf{q},\omega) \right) \mathrm{d} \omega}{ \sum_\mathbf{q}\int  e^{-\beta\omega} \left(\hat A_{\uparrow\uparrow}(\mathbf{q},\omega)+\hat A_{\downarrow\downarrow}(\mathbf{q},\omega) \right)\mathrm{d} \omega}.
    \label{eq:mag}
\end{equation}
Note that this expression is independent of $\mu_\mathrm{imp}$.  

\begin{figure}
    \centering
	\includegraphics[width=\columnwidth]{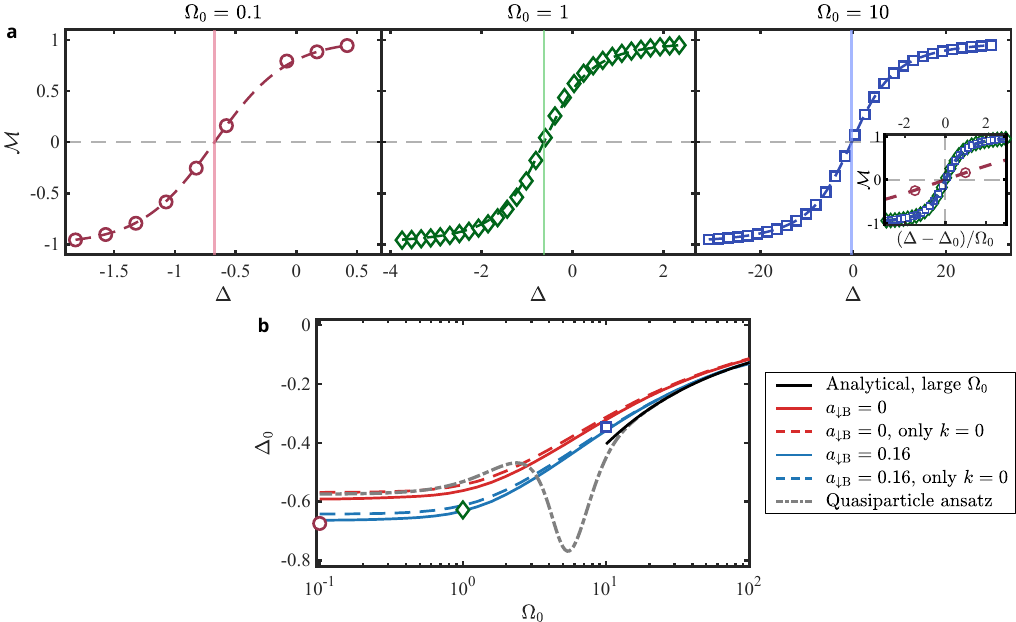}
	\caption{\textbf{Theoretical steady-state magnetization and zero crossing.} (\textbf{a}) Steady-state magnetization obtained from the $T$-matrix approximation Eq.~\eqref{eq:mag} (empty colored symbols) together with a fit using Eq.~\eqref{eq:mag_pheno} (colored dashed lines). Vertical lines mark the fitted $\Delta_0$. Inset: Scaling collapse of the steady-state spectrum along with the rescaled fit (colored dashed lines). The inverse temperature is $\beta^{-1}=0.3$. 
     (\textbf{b}) Zero crossing evaluated within various approximations. The red and blue lines are evaluated using the in-medium $T$ matrix, while the gray dash-dotted line is evaluated from the quasiparticle ansatz Eq.~\eqref{eq:MQP_ex}. For the solid (dashed) lines, we evaluate the full momentum integration (only the zero-momentum part) of Eq.~\eqref{eq:mag}. The red (blue) lines refer to $a_{\downarrow\B}=0$ ($a_{\downarrow\B}=0.16$). The black solid line is the exact result in the large $\Omega_0$ limit (see text). The empty symbols are the extracted $\Delta_0$ from the panel (\textbf{a}). \label{Fig:S7}}
\end{figure}

In Fig.~\ref{Fig:S7}a we show magnetization spectra obtained from Eq.~\eqref{eq:mag} using the $T$-matrix spectral functions (colored symbols), along with fits to Eq.~\eqref{eq:mag_pheno} where $\Delta_0$ -- the zero crossing as defined in the main text -- is the only fit parameter (black dashed lines). These fits describe the numerical results well. In the inset of Fig.~\ref{Fig:S7}a we show the rescaled magnetization, similar to the inset of Fig. 2b in the main text. These rescaled magnetizations also collapse into a universal profile when $\beta\Omega_0\gg1$. Finite temperature corrections lead to a breakdown of the scaling collapse, which is visible for the $\Omega_0=0.1$ plot (for which $\beta\Omega_0=0.33$).

\subsubsection{Magnetization from a quasiparticle ansatz}

To get a qualitative understanding of the many-body magnetization in terms of the quasiparticle properties, we calculate the magnetization Eq.~\eqref{eq:mag} using the quasiparticle spectral functions Eqs.~\eqref{eq:specfuncupup}-\eqref{eq:specfunc_22}.  
To make the calculation tractable, we slightly oversimplify by assuming that $m_\uparrow^* = m_\downarrow^* = m$ (a reasonable assumption since $m_\uparrow^*\approx m_\downarrow^*$ for both small and large $\Omega_0$), such that the momentum dependence in Eq.~\eqref{eq:mag} drops out by substituting $\omega$ for $\omega-\epsilon_\mathbf{q}$. We obtain
\begin{equation}
    \mathcal{M}_{\mathrm{QP}}=\frac{\int  e^{-\beta\omega} \left(\hat A^{\mathrm{QP}}_{\uparrow\uparrow}(\omega)-\hat A^{\mathrm{QP}}_{\downarrow\downarrow}(\omega) \right) \mathrm{d} \omega}{\int  e^{-\beta\omega} \left(\hat A^{\mathrm{QP}}_{\uparrow\uparrow}(\omega)+\hat A^{\mathrm{QP}}_{\downarrow\downarrow}(\omega) \right)\mathrm{d} \omega},
    \label{eq:magqp}
\end{equation} 
where we define $\hat A^{\mathrm{QP}} _{\alpha\alpha}(\omega)\equiv \hat A^{\mathrm{QP}}_{\alpha\alpha}(\mathbf{q},\omega+\epsilon_\mathbf{q})=\hat A^{\mathrm{QP}}_{\alpha\alpha}(\mathbf{q}=0,\omega)$.

However, the integrals in Eq.~\eqref{eq:magqp} do not converge since the integrands diverge at $\omega\rightarrow-\infty$. This is an artefact of the quasiparticle ansatz because of the assumption that the self-energy is  $\omega$-independent (see the discussion in Sec.~\ref{section: Qp properties}). To overcome this problem, we replace the Lorentzians in Eq.~\eqref{eq:specfuncupup}-\eqref{eq:specfunc_22} by Dirac deltas peaked at the same frequencies. Furthermore, we assume $\Gamma\tilde \Delta\ll \Omega_0$ and $\Gamma/2< \sqrt{Z}\Omega_0$ such that $\OmegaR^\mathrm{i}\approx 0$ and $\OmegaR=\sqrt{Z\Omega_0^2+(\Delta-E_\mathrm{p})^2}$. Within these approximations, we find an analytic expression for the magnetization:
 
\begin{equation}
	\mathcal{M}_{\mathrm{QP}} = \frac{(\Delta-E_\mathrm{p}) \tanh\left(\frac{\beta \OmegaR}{2}\right)+\frac{Z-1}{Z+1}\OmegaR}{\OmegaR+\frac{Z-1}{Z+1}(\Delta-E_\mathrm{p})\tanh\left(\frac{\beta \OmegaR}{2}\right)}.
	\label{eq:MQP_ex}
\end{equation}
Finally, if $Z\approx 1$, we find
\begin{equation}
	\mathcal{M}_{\mathrm{QP}} \approx \frac{\Delta-E_\mathrm{p}}{\OmegaR}\tanh\left(\frac{\beta \OmegaR}{2}\right),
\end{equation}
a magnetization of the same form as that of the dressed spin $1/2$ in thermal equilibrium Eq.~\eqref{eq:mag_pheno}, for which the resonance frequency is shifted by $E_\mathrm{p}$, \emph{i.e.}, $\Delta_0=E_\mathrm{p}$.

For $Z\neq 1$, we find the following corrections:
\begin{equation}
    \Delta_0=
    \begin{cases}
        E_\mathrm{p} - \frac{1}{2}\Omega_0(Z-1)&\Omega_0\gg1\\
        E_\mathrm{p} + \frac{2}{\beta} \mathrm{arctanh}\left(\frac{1-Z}{Z+1}\right)&\Omega_0\ll 1.
    \end{cases}
    \label{eq:largeOMZC}
\end{equation}

\subsubsection{Zero crossing: numerical evaluation}

In Fig.~\ref{Fig:S7}b, we show the zero crossing evaluated with different approximations. 
 First, we assess the importance of the sum over $\mathbf{q}$ in Eq.~\eqref{eq:mag} by evaluating the magnetization taking only the $\mathbf{q}=\mathbf{0}$ term into account (dashed colored lines) or the full sum (solid colored lines). Apart from a shift at small $\Omega_0$, these predictions overlap at large $\Omega_0$, in line with the expectation that finite-temperature effects are less important at large $\Omega_0$.

 We estimate the effect of the (weak) $\ket{\downarrow}$-$\ket{\B}$ interaction by showing the results for $a_{\downarrow\B}=0.16$ (blue lines) and for $a_{\downarrow\B}=0$ (red lines). We observe a shift of $\Delta_0$ of $\approx 0.07$ for small $\Omega_0$, which corresponds to the energy of the repulsive polaron formed in the $\ket{\downarrow}$-$\ket{\B}$ mixture (Eq.~\eqref{eq: E+}). This shift is smaller as $\Omega_0$ increases. 

We extract the zero crossing from the quasiparticle-ansatz magnetization Eq.~\eqref{eq:MQP_ex}, using Eq.~\eqref{eq:QP_props}. In the range $1\lesssim\Omega_0\lesssim10$, the quasiparticle ansatz predicts an implausible non-monotonic behavior (gray dash-dotted line in Fig.~\ref{Fig:S7}), that is absent from the experimental data and the $T$-matrix results. We attribute this behavior to the fact that the line shapes -- especially of the lower branch due to the factor $e^{-\beta\omega}$ in Eq.~\eqref{eq:magqp} -- and peak heights of the spectral functions are poorly captured by the quasiparticle ansatz. In particular, between $\Omega_0\approx 3$ and $\Omega_0\approx 10$, $\Delta_0$ deviates most significantly from the $T$-matrix predictions, which might -- again -- be related to the merging of the upper-branch polaron with the incoherent background (see Fig.~\ref{Fig:S5}b and discussions in Sec.~\ref{compsrison of spectral functions}). This shows that the assumption of using Dirac deltas to evaluate the magnetization is invalid in this regime. On the other hand, for small and large $\Omega_0$, the quasiparticle approximation reproduces the $T$-matrix results better (up to a small shift at small $\Omega_0$), which is in line with the observation in Fig.~\ref{Fig:S5}b that the $T$-matrix spectral functions are sharply peaked, and thus the approximation of replacing the Lorentzians in the quasiparticle spectral functions with Dirac deltas is reasonable in this regime. 

Finally, we find that all approximations converge towards the large drive limit (black solid line, $\mathrm{Max}\left(1/a_{\uparrow \mathrm{B}},1\right)\ll \sqrt{\Omega_0}/2 \ll  1/a_{\downarrow \mathrm{B}}$), obtained by inserting Eqs.~\eqref{eq:largeOmEp}-\eqref{eq:largeOmZ} into the $\Omega_0\gg1$ limit of Eq.~\eqref{eq:largeOMZC}.

\subsubsection{Zero crossing: comparison to the experiment}

\begin{figure}	
\includegraphics{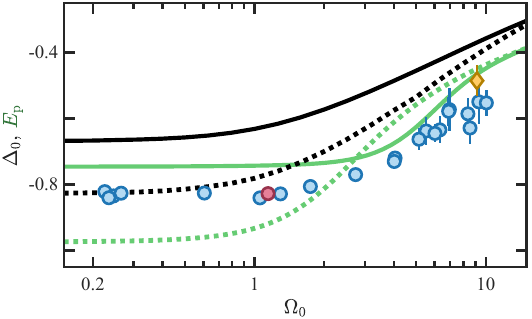}
	\caption{\textbf{Zero crossing and drive-dressed polaron energy as a function of Rabi frequency.} The black solid (dotted) line is the theoretical prediction for $\Delta_0$ using the in-medium (two-body) $T$ matrix. The green solid (dotted) line is the theoretical prediction for $\Ep$ using the in-medium (two-body) $T$ matrix.  All theoretical results are evaluated at $a_{\downarrow\B}=0.16$ and $\beta^{-1}=0.25$. The data shown here is the same as in Fig.~2b of the main text but rescaled with $\EF$ rather than the value in the limit of $ \Omega_0\rightarrow 0$. The red circle and yellow diamond refer to the measurements in Fig.~2a in the main text. }\label{Fig:S8}
\end{figure}

In Fig.~\ref{Fig:S8}, we show the experimental measurement of $\Delta_0$ along with the in-medium (black solid line) and two-body (black dotted line) $T$-matrix predictions, as well as the drive-dressed polaron energy $E_\mathrm{p}$ (green solid line), as in Fig.~2b of the main text, but rescaled to $\EF$ instead of the reference quantities. The reference quantities chosen in Fig.~2b, $\Delta_0^{\mathrm{ref}}$ and $\Ep^{\mathrm{ref}}$, correspond to the plateau value of $\Omega_0\rightarrow 0$ of the black and green solid lines.

\subsection{Connecting Rabi oscillations to equilibrium properties\label{sct:Rabifromspec}}

Here, we use the arguments of Refs.~\cite{Adlong2020S,Hu2022S} to connect Rabi oscillations to the spectral functions, emphasizing the approximations involved. We assume that the initial state $\rho$ is a product state of a (single-component) non-interacting Fermi gas and a single impurity in the state $\ket{\downarrow}$ with zero momentum. This assumption is reasonable as the impurity in $\ket{\downarrow}$ is barely interacting with the bath and we typically work at very low impurity concentrations (see Sec.~\ref{sec:finiteconc}). The initial state is thus $\rho =  c^\dagger_\downarrow  \rho_{\mathrm{FS}}  c_\downarrow$, where $\rho_\mathrm{FS}$ is the density matrix of the non-interacting Fermi gas and $c_\downarrow$ ($c_\downarrow^\dagger$) is the annihilation (creation) operator for the zero-momentum impurity in state $\ket{\downarrow}$. The time-dependent magnetization is thus
\begin{align}
    \mathcal{M}(t) &= \sum_\mathbf{q}\Tr\left[ \rho  c^\dagger_{\mathbf{q}\uparrow} (t)  c_{\mathbf{q}\uparrow} (t)  \right]-\sum_\mathbf{q}\Tr\left[ \rho  c^\dagger_{\mathbf{q}\downarrow} (t)  c_{\mathbf{q}\downarrow} (t)  \right] \\
    &= \sum_\mathbf{q}\Tr\left[ \rho_{\mathrm{FS}}  c_\downarrow c^\dagger_{\mathbf{q}\uparrow} (t)  c_{\mathbf{q}\uparrow} (t)  c_\downarrow^\dagger \right]-\sum_\mathbf{q}\Tr\left[ \rho_{\mathrm{FS}}  c_\downarrow c^\dagger_{\mathbf{q}\downarrow} (t)  c_{\mathbf{q}\downarrow} (t)  c_\downarrow^\dagger \right],
\end{align}
where $c_{\mathbf{q}\alpha}(t)=e^{iHt}c_{\mathbf{q}\alpha}e^{-iHt}$ and $c_{\mathbf{q}\alpha}^\dagger(t)=e^{iHt}c_{\mathbf{q}\alpha}^\dagger e^{-iHt}$ follow the Heisenberg equation of motion with the Hamiltonian Eq.~\eqref{eq:Hamiltonian}. To calculate this four-body correlation function, we use a mean-field-type approximation, in which higher-order correlation functions factorize~\cite{Adlong2020S,Hu2022S}:
\begin{align}
    \mathcal{M}(t)&\approx \Tr\left[ \rho_{\mathrm{FS}}  c_\downarrow c^\dagger_{\uparrow} (t)\right]\Tr\left[ \rho_{\mathrm{FS}} c_{\uparrow} (t) c_\downarrow^\dagger \right]-\Tr\left[ \rho_{\mathrm{FS}}  c_\downarrow c^\dagger_{\downarrow} (t)\right]\Tr\left[ \rho_{\mathrm{FS}} c_{\downarrow} (t)  c_\downarrow^\dagger \right]\\
    &=\bigg| \int e^{i\omega t} \hat A_{\uparrow\downarrow} (\omega,\mathbf{k}=\mathbf{0}) \mathrm{d} \omega \bigg|^2-\bigg| \int e^{i\omega t} \hat A_{\downarrow\downarrow} (\omega,\mathbf{k}=\mathbf{0}) \mathrm{d} \omega \bigg|^2.
\end{align}
Inserting the quasiparticle spectral function Eqs.~\eqref{eq:specfunc_22}-\eqref{eq:specfunc_12} for resonant ($\tilde{\Delta}=0$) and strongly underdamped ($Z\Omega_0^2\gg\Gamma^2$) Rabi oscillations, we find
\begin{equation}
    \mathcal{M}(t)=e^{-\Gamma t/2} \left( \frac{Z-1}{2}-\frac{Z+1}{2}\cos(\OmegaR t)\right).
\end{equation}
This expression obeys the expected asymptotic behaviors, $\mathcal{M}(t=0)=-1$ and $\mathcal{M}(t\rightarrow\infty)=0$. The oscillation frequency and decay rate are $\OmegaR=\sqrt{Z}\Omega_0$ and $\Gamma/2$, respectively. However, we also find that this mean-field approximation does not conserve the impurity number as
\begin{align}
    \Tr\left[ c_\downarrow^\dagger\rho_{\mathrm{FS}}c_\downarrow  \left(c^\dagger_\downarrow (t)c_\downarrow (t)+c^\dagger_\uparrow (t)c_\uparrow (t)\right)\right]&\approx
    \Tr\left[ \rho_{\mathrm{FS}}  c_\downarrow c^\dagger_{\uparrow} (t)\right]\Tr\left[ \rho_{\mathrm{FS}} c_{\uparrow} (t)  c_\downarrow^\dagger \right]+\Tr\left[ \rho_{\mathrm{FS}}  c_\downarrow c^\dagger_{\downarrow} (t)\right]\Tr\left[ \rho_{\mathrm{FS}} c_{\downarrow} (t)  c_\downarrow^\dagger \right]\\
    &\approx e^{-\Gamma t/2} \left( \frac{Z+1}{2}-\frac{Z-1}{2}\cos(\OmegaR t)\right).
\end{align}
Hence, the mean-field approximation is ultimately unsatisfactory, and higher-order corrections are needed. It is thus remarkable that despite this issue, the decay rate and the renormalized Rabi frequency agree with the experimental measurements (see Fig.~3 in the main text).

\section{Spectroscopic probe of the Fermi energy with the repulsive polaron energy}

We calibrate our Fermi energy using the spectroscopic method introduced in Ref.~\cite{Scazza2017S}. To this end, we prepare a spin-imbalanced gas with a concentration $x\approx 0.15$ (as described in the main text) and ramp the magnetic field to final values between $\unit[630-650]{G}$. At those fields, a well-defined repulsive polaron exists for the $\ket{\uparrow}$ state ($k_\mathrm{F}a_{\uparrow \mathrm{B}} < 0.45$). The repulsive polaron energy is given in the weak coupling limit by~\cite{massignan2011repulsiveS}
\begin{equation}\label{eq: E+}
    \frac{E_+}{E_\mathrm{F}}=\frac{4}{3\pi}k_\mathrm{F}a+\frac{2}{\pi^2}(k_\mathrm{F}a)^2+\mathcal{O}[(k_\mathrm{F}a)^3].
\end{equation}

\begin{figure}
	\includegraphics{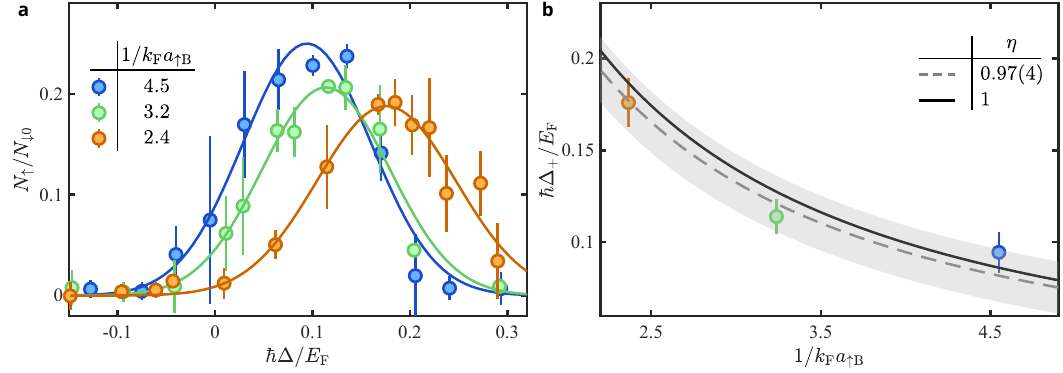}
	\caption{\textbf{Calibration of $E_{\mathrm{F}}$ by repulsive polaron spectroscopy.} ($\mathbf{a}$) Linear-response spectroscopy of the weakly repulsive polaron. We use a \unit[1]{ms} rf pulse with a Rabi frequency $\Omega_0\approx2\pi\times \unit[150]{Hz}$. The spectroscopic shift $\Delta_+$ is extracted by a Gaussian fit (solid lines). The error bars are the s.e.m of the measurement. Colors represent different $k_\mathrm{F}a_{\uparrow \mathrm{B}}$ (see legends). $N_{\downarrow0}$ ($N_{\uparrow}$) is the initial (transferred) impurity atom numbers in state $\ket{\downarrow}$ ($\ket{\uparrow}$). ($\mathbf{b}$) Extraction of the effective Fermi energy. The dashed line is a fit to Eq.~\eqref{eqn:RepulsivePolaronShift} with $\epsilon_\F=\eta E_\F$. The solid line is the prediction from Eq.~\eqref{eqn:RepulsivePolaronShift} with $\eta=1$. The gray band is the uncertainty of the fit. Error bars are a combination of s.e.m and uncertainty of the fit (here, $\EF/h\approx\unit[5.7]{kHz}$)}\label{Fig:S9}
\end{figure}

Taking into account the nonzero $\kF a_{\downarrow\B}$, we expect the spectroscopic shift $\Delta_+$ to be
\begin{equation}\label{eqn:RepulsivePolaronShift}
    \frac{\hbar\Delta_+}{E_\mathrm{F}}=\frac{4}{3\pi}k_\mathrm{F}(a_{\uparrow \mathrm{B}}-a_{\downarrow \mathrm{B}})+\frac{2}{\pi^2}k_\mathrm{F}^2(a_{\uparrow \mathrm{B}}^2-a_{\downarrow \mathrm{B}}^2).
\end{equation}
In Fig.~\ref{Fig:S9}a we show spectra measured at three different fields, along with (Gaussian) fits to extract $\Delta_+$. To model the uncertainty of this calibration due to the atom number counting and the volume determination, we define the effective Fermi energy $\epsilon_\mathrm{F}=\eta E_\mathrm{F}$, calibrated from the spectroscopic shift as a function of $E_\mathrm{F}$, that is determined from our atom numbers and extracted box volume. From the fitting of $\Delta_+$ (Fig. \ref{Fig:S9}b), we obtain $\eta = 0.97(4)$, which validates our calibration of $E_\mathrm{F}$ and the spatial homogeneity of the box.

\section{Effects of finite impurity concentration}
\label{sec:finiteconc}

\begin{figure}
	\includegraphics{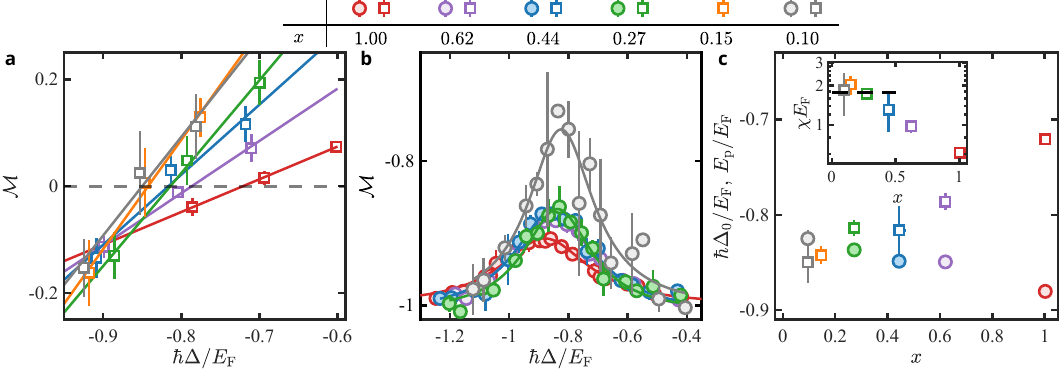}
	\caption{\textbf{Effect of finite impurity concentration.} (\textbf{a}) Steady-state spectroscopy and (\textbf{b}) linear-response spectroscopy for various concentrations (see legends). $\Omega_0/(2\pi)\approx \unit[1]{kHz}$ in (\textbf{a}) and \unit[155]{Hz} in (\textbf{b}), except for $x\approx0.1$ in (\textbf{b}), where we use $\Omega_0/(2\pi)\approx\unit[220]{Hz}$ to improve the signal. (\textbf{c}) $\Delta_0$ (empty colored squares) and $E_\mathrm{p}$ (filled colored circles) extracted from (\textbf{a}) and (\textbf{b}). Inset: $\chi$ extracted from (\textbf{a}). The black dashed line is the prediction from the finite-temperature magnetization profile Eq.~\eqref{eq:mag_pheno}. Error bars in (\textbf{a}) and (\textbf{b}) are the s.e.m of the measured numbers, and in (\textbf{c}) are the uncertainties of the fit.}\label{Fig:S10}
\end{figure}

Here, we estimate the effect of the finite impurity concentration on the measurement of $\Delta_0$ and $\Ep$. As shown in Fig.~\ref{Fig:S10}a and Fig.~\ref{Fig:S10}b, we measure $\Delta_0$ and $E_\mathrm{p}$ of an imbalanced Fermi gas for various minority concentrations $x$, by steady-state- and linear-response spectroscopy, respectively. They are shown in Fig.~\ref{Fig:S10}c, where we find at most a difference of $13\%$ (resp. $6\%$) for $\Delta_0$ (resp. $\Ep$) with the values reported in the main text in the range of $0.1\lesssim x\lesssim 1.0$. 
In particular, we find that $\Delta_0$ and $E_\mathrm{p}$ are affected differently by the finite impurity concentration.

\section{Coherence of Rabi oscillations}

To calibrate $\Omega_0$, we prepare a spin-polarized Fermi gas in the state $\ket{\downarrow}$ and apply an rf pulse of duration $t$, tuned on resonance with the $\ket{\downarrow}$-$\ket{\uparrow}$ transition. The resulting magnetization is fitted with $\mathcal{M}(t)=-\cos(\Omega_0 t)\exp(-\Gamma_\mathrm{bg} t/2)$, where $\Gamma_\mathrm{bg}$ is a (slow) background decoherence rate. For the measurement of $\OmegaR$ and $\Gamma$, we prepare instead a highly imbalanced mixture of states $\ket{\downarrow}$-$\ket{\mathrm{\mathrm{B}}}$, and apply an rf pulse of detuning $\Delta_0$. The oscillations are fitted with $\mathcal{M}(t)=-\cos(\OmegaR t)\exp(-\Gamma t/2)$. 

\begin{figure}
	\includegraphics{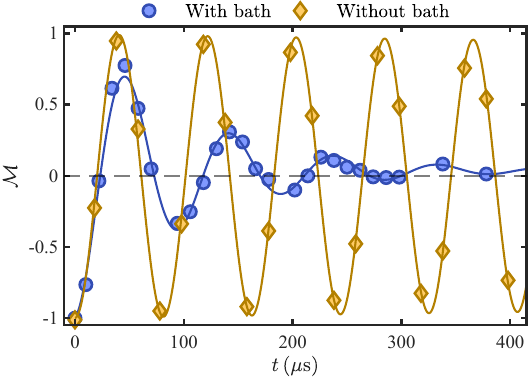}
	\caption{\textbf{Rabi oscillations with and without the bath.} The blue circles (yellow diamonds) are the measurements of the Rabi oscillations of the impurity internal states with (without) the bath $\ket{\B}$. The colored solid lines are fits, see text.}
\label{Fig:S11}
\end{figure}

A typical measurement is shown in Fig.~\ref{Fig:S11}, where the spin-polarized (resp. highly imbalanced) data corresponds to the yellow diamonds (resp. blue circles). The colored solid lines are the fits. The fitted Rabi frequency is $\Omega_0/(2\pi)=$\unit[12.30(1)]{kHz} (resp. $\OmegaR/(2\pi)=$\unit[10.46(7)]{kHz}), and the decay rate is $\Gamma_\mathrm{bg}/(2\pi)=$\unit[0.03(1)]{kHz} (resp. $\Gamma/(2\pi)=$\unit[3.1(1)]{kHz}). In particular, the typical $\Gamma$ is about two orders of magnitude larger than $\Gamma_\mathrm{bg}$. The decay of the Rabi oscillations in the imbalanced case is thus not dominated by technical sources.

\begin{thebibliography}{10}
\bibitem{footnote1}
Our measurement would be equivalent to linear-response injection spectroscopy
  if performed in the regime of weak drive ($\hbar\Omega_0/\EF\ll 1$) and short
  times (for which $\mathcal{M}\approx -1$).
\bibitem{footnote}
We note that for $\Omega_0\rightarrow 0$, the theoretical $\Delta_0$ and
  $E_\textrm{p}$ differ at $T>0$; this is not only a property of the $T$-matrix
  approach, but also of any quasiparticle ansatz (see methods and
  SM~\cite{supp}).
\end{thebibliography}
\end{document}